\documentclass[prd,twocolumn,amsmath,amssymb,aps,nofootinbib]{revtex4-2}

\usepackage{aas_macros}
\usepackage{bm}
\usepackage{gensymb}
\usepackage{epsfig}
\usepackage{url}
\usepackage{hyperref}

\usepackage[normalem]{ulem} 

\usepackage{latexsym}
\usepackage{epsfig}
\usepackage{amsmath}
\usepackage{amssymb}
\usepackage{wasysym}
\usepackage{graphicx}
\usepackage{dcolumn}
\usepackage{verbatim}
\usepackage{enumerate,mdwlist}
\usepackage[titletoc]{appendix}
\usepackage{amsfonts}
\usepackage{fontawesome}
\usepackage{fancyvrb} 
\usepackage{tikz} 
\usetikzlibrary{calc}
\usepackage[export]{adjustbox}

\usepackage{booktabs}

\usepackage{multirow}

\usepackage{pifont}  

\usepackage[normalem]{ulem}

\newcommand{\ba}{\begin{eqnarray}}
\newcommand{\ea}{\end{eqnarray}}
\newcommand{\be}{\begin{equation}}
\newcommand{\ee}{\end{equation}}

\newcommand{\Msun}{\mathrm{M}_{\odot}} 

\definecolor{grey}{rgb}{0.4,0.4,0.4}
\definecolor{dullmagenta}{rgb}{0.4,0,0.4}
\definecolor{darkblue}{rgb}{0,0,0.4}
\definecolor{midblue}{rgb}{0,0,0.5}
\definecolor{midred}{rgb}{0.5,0,0}
\definecolor{orange}{rgb}{1,0.5,0}
\definecolor{lightbrown}{rgb}{0.75,0.5,0.25}
\definecolor{tan}{cmyk}{0.14,0.42,0.56,0}
\definecolor{djunglegreen}{cmyk}{0.99,0,0.52,0}
\definecolor{lightgreen}{rgb}{0,1,0}
\definecolor{olivegreen}{cmyk}{0.64,0,0.95,0.40}
\definecolor{midgreen}{rgb}{0.0,0.675,0.0}
\definecolor{darkgreen}{rgb}{0,0.5,0}
\definecolor{ceruleanblue}{rgb}{0.0, 0.2, 0.7}
\definecolor{burgunday}{rgb}{0.5, 0.0, 0.13}
\definecolor{hvred}{RGB}{186,12,47}

\hypersetup{
    colorlinks=true,
    linkcolor=ceruleanblue,
    filecolor=ceruleanblue,      
    urlcolor=midblue,
    citecolor=burgundy,
}

\newcommand{\change}[1]{\textcolor{black}{#1}}

\usepackage[all]{xy} 
\usepackage{amsfonts}

\makeatletter
\def\l@subsubsection#1#2{}
\makeatother

\newcommand{\qpls}{{QPLS}}
\newcommand{\Qplsname}{{Quasi-periodic lensing of starlight}}
\newcommand{\qplsname}{{quasi-periodic lensing of starlight}}

\usepackage{xcolor}
\definecolor{burgundy}{rgb}{0.5, 0.0, 0.13}
\begin{document}


\title{Black holes as telescopes: \\Discovering supermassive binaries through quasi-periodic lensed starlight}



\author{Hanxi Wang}
\email[Email: ]{hanxi.wang@physics.ox.ac.uk}
\affiliation{Department of Physics, Astrophysics, University of Oxford, Denys Wilkinson Building, Keble Road, Oxford, OX1 3RH, UK}

\author{Miguel Zumalac\'arregui}
 \email[Email: ]{miguel.zumalacarregui@aei.mpg.de}
 \affiliation{Max Planck Institute for Gravitational Physics (Albert Einstein Institute) \\
 Am Mühlenberg 1, D-14476 Potsdam-Golm, Germany}

\author{Bence Kocsis}
\email[Email: ]{bence.kocsis@physics.ox.ac.uk}
\affiliation{Rudolf Peierls Centre for Theoretical Physics, University of Oxford, Clarendon Laboratory, Parks Road, Oxford, OX1 3PU, UK}
\affiliation{St Hugh’s College, University of Oxford, St Margaret’s Rd, Oxford, OX2 6LE, UK}

\begin{abstract}
Supermassive black hole (SMBH) binary systems are an unavoidable outcome of galaxy mergers. Their dynamics encode valuable information about their formation and growth, the composition of their host galactic nuclei, the evolution of galaxies, and the nature of gravity. Many SMBH binaries with separations pc-kpc have been found, but closer (sub-parsec) binaries remain to be confirmed. Identifying these systems may elucidate how binaries evolve past the ``final parsec'' until gravitational radiation drives them to coalescence. Methods to discover and characterize SMBH binaries can shed light on these important questions and potentially open new multi-messenger channels. Here we show that SMBH binaries in non-active galactic nuclei can be identified and characterized by the gravitational lensing of individual bright stars, located behind them in the host galaxy. The rotation of `caustics' -- regions where sources are hugely magnified due to the SMBH binary's orbit and inspiral--  leads to \qplsname{} (\qpls). The extreme lensing magnification of individual bright stars produces a significant variation in the host galaxies' luminosity;  their lightcurve traces the orbit of the SMBH binary and its evolution, analogous to the waveforms recorded by gravitational-wave (GW) detectors. \qpls{} probes the population of sources observable by pulsar timing arrays and space detectors (LISA, TianQin), offering advance warning triggers for merging SMBHs for coincident or follow-up GW detections. SMBH population models predict $1-50\; [190-5,000] \left({n_\star}/{\rm pc}^{-3}\right)$ \qpls{} binaries with period less than $10\; [40]$ yr with comparable masses and redshift $z<0.3$, where $n_\star$ is the stellar number density. Additionally, stellar and orbital motion will lead to frequent instances of single/double flares caused by SMBHBs with longer periods. This novel signature can be searched for in a wealth of existing and upcoming time-domain photometric data: identifying quasi-periodic variability in galactic lightcurves will reveal an ensemble of binary systems and illuminate outstanding questions around them.
\end{abstract}

\date{\today}




\maketitle


\paragraph*{\bf Introduction.} 

Astronomical observations demonstrate that most galaxies harbor a supermassive black hole (SMBH) at their center. As galaxies merge, their central black holes encounter each other, forming bound systems. The evolution of these systems is driven by interactions with the surrounding star cluster, gaseous drag, and the emission of gravitational radiation.~\cite{Begelman:1980vb,Escala2004,Volonteri_2010_review}. Identifying SMBH binary (SMBHB) systems will elucidate interactions between central black holes and galactic nuclei~\cite{Milosavljevic:2002bn}~\cite[p.~81, Fig.~17]{LISA:2022yao}, their role in galaxy formation and evolution \cite{Volonteri2003,Kormendy2013}, provide new tests of gravity and fundamental physics~\cite{Alonso-Alvarez:2024gdz,Kocsis+2008}. Many SMBHBs with parsec (pc) to kiloparsec (kpc) separation have been identified~\cite{DeRosa:2019myq}. Candidates for tighter binaries with sub-parsec separation need to be confirmed~\cite{DOrazio:2023rvl}. 

Tight, sub-pc SMBHBs evolve via gravitational wave (GW) emission, opening the possibility of multi-messenger observations if coincident GW and electromagnetic (EM) emission is identified~\cite{Kocsis+2008,Bogdanovic:2021aav,Charisi:2021dwc}.
Building evidence for a nanoHz GW background using pulsar timing arrays (PTA)~\cite{NANOGrav:2023gor,EPTA:2023fyk,Reardon:2023gzh,Xu:2023wog} sheds light on the population of binaries with mass $\gtrsim 10^8\Msun$~\cite{Ellis:2023dgf,Sato-Polito:2025ivz}. 
Space-borne GW detectors in the mHz band, LISA~\cite{LISA:2017pwj,LISA:2024hlh} and TianQin~\cite{TianQin:2015yph,Luo:2025sos}, will record mergers of binaries $10^5-10^8\Msun$ and open new windows into astrophysics and cosmology~\cite{LISA:2022yao,LISACosmologyWorkingGroup:2022jok}. 

Identifying tight SMBHBs with EM observations is a challenging task. Most methods require the SMBHB to be embedded in gas and shine as a bright active galactic nucleus (AGN), and rely on detecting modulations of the emission of the accreted gas, where the binary motion imprints regular features modulated by the orbital period due to e.g. gravitational torques, accretion variability, or relativistic Doppler boost~\cite{Haiman_Kocsis_Menou2009,DOrazio:2015jcb,Charisi:2018qyp,Xin:2019dkw,Song:2019lxb}. 
SMBHBs can also be discovered by gravitational lensing, the deflection and magnification of light signals by gravitational fields~\cite{Schneider:1992,Bozza:2010xqn}. 
One possibility is the lensing of a black hole's accretion disk by its companion, potentially leading to periodic flares~\cite{DOrazio:2017ssb,Davelaar:2021eoi,Kelley+2021,Krauth+2024}, or the lensing of blazar jets by an intervening intermediate mass black hole binary \cite{Krol:2022wpj}. Other studies consider the effect of SMBHBs on central images in galaxy-galaxy lenses, whose faint modulations may be detectable~\cite{Li:2011js,Hezaveh:2015oya}.
To the best of our knowledge the lensing effect of SMBHBs on individual stars in the host galaxy remains unexplored.

Here we demonstrate how SMBHBs can act as telescopes, magnifying individual stars' brightness comparable to their host galaxy: the orbital motion generates \qplsname{} (\qpls), a characteristic time-dependent photometric lightcurve revealing the SMBHB's presence and dynamics. This method bears resemblance to microlensing by binary stars or by exoplanets, where the light of a star in the Milky-Way is lensed by a star-planet binary system crossing the line-of-sight: this imprint reveals the binary properties~\cite{Mao_Paczynski91,Gould_Loeb1992,Gaudi+2012}.
For \qpls{}, the source magnification can be much higher than in exoplanet searches, and even surpass highly-amplified individual stars found in giant arcs of galaxy clusters~\cite{Miralda-Escude1991,Kelly:2017fps,Diego:2018fzr}. Note that this method does not require gas to be present around the SMBHBs, which may be an advantage given that more than $90\%$ of galaxies do not produce an AGN \cite{Kauffmann+2003}.

The likelihood of lensing by an inspiraling SMBHB is greatly increased compared to a static lens. The changing position and separation of the binary increases the solid angle of source directions affected by lensing. The caustic curve of the SMBHB lens rotates and slowly shrinks with the inspiraling binary's orbit, it crosses the image of the star repeatedly, and creates a quasi-periodic light curve. The timing and shape of this \qpls{} lightcurve carries information on the mass, period, inclination, and eccentricity of the SMBHB.

\begin{figure*}
    \centering
    \includegraphics[width=0.99\linewidth]{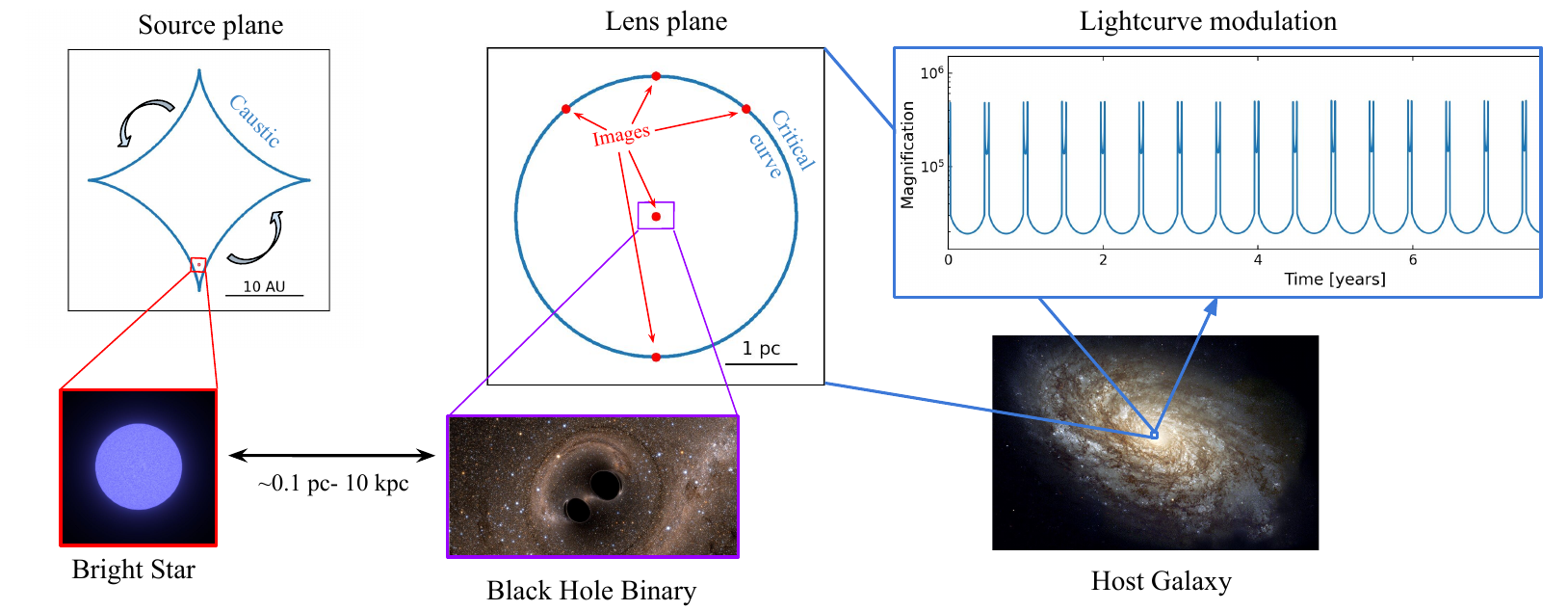}
    \caption{\Qplsname: a bright star (left) is highly magnified by a binary supermassive black hole binary (center). The caustic rotates as the binary orbits, producing a bright quasi-periodic signal at the central region of the host galaxy (right). The SMBHB in the diagram has total mass of $10^{10} \Msun$ and an initial period of 2 years. The distance between the lens and source plane is 1 kpc, the star has a size of $10R_{\odot}$. The red dots in the lens plane mark the positions of the images. There are 5(3) images when the source is inside (outside) the caustic curves, out of which 4(2) images are near the critical curve hence highly magnified. In addition to the diamond-shaped caustic, two triangle-shaped caustics form at about 230 pc from the center of mass (not shown). Credit: NASA/ESA, SXS/AEI, ESO~\cite{composite-figures}.
    } 
    \label{fig:master_diagram}
\end{figure*}

\paragraph*{\bf High Magnification by Binary Lenses.}

We model the SMBHB as a two-point-mass lens \cite{schneider:1986}. The lenses have mass $M_1$ and $M_2$ and we approximate them as projected to a common lens plane (perpendicular to the line of sight) at an angular diameter distance $D_{\rm L}$ from the observer. The source is a star in the same galaxy as the SMBHB, whose distance to the observer ($D_S$) and to the lens plane ($D_{\rm LS}$) satisfy $D_{\rm LS}\ll D_{\rm L}\sim D_{S}$. The Einstein radius of this system corresponds to a distance
\begin{equation}
    \label{eq:ein_rad}
    \xi_{0} =  2  \sqrt{R_{\rm g} \frac{D_LD_{\rm LS}}{D_S}} \approx
    2\sqrt{R_{\rm g}
    D_{\rm LS}} =1.38\text{pc}\, M_{10}^{1/2}{D_{\rm kpc}^{1/2}},
\end{equation}
where $R_{\rm g} = GM/c^2$, $M=M_1+M_2$ is the total mass of the binary, $M_{10}=M/(10^{10}\Msun)$, and $D_{\rm kpc}=D_{\rm LS}/\rm kpc$.
The lens equations is expressed in units of the Einstein radius as
\begin{align}
\label{eq:leqn}
\bm{y} &=\bm{x}- m_1 \frac{\bm{x}-\bm{x}_{m_1}}{|\bm{x}-\bm{x}_{m_1}|^2}-m_2 \frac{\bm{x}-\bm{x}_{m_2}}{|\bm{x}-\bm{x}_{m_2}|^2},
\end{align}
where $(D_{\rm S}/D_{\rm L})\xi_0\bm{y}$ is the source position in the source plane, $\xi_0\bm{x}$ is the position of the image in the lens plane,  $\xi_0\bm{x_{m_1}}$ and $\xi_0\bm{x_{m_2}}$ are the positions of the two point-mass lenses in the lens plane, and $m_1$ and $m_2$ are given by $m_{1,2} = M_{1,2}/(M_1+M_2)$.\footnote{We will refer to source frame quantities, unless otherwise stated.}

The magnification $\mu$ of an image is computed from the Jacobian of the lens equation, $\mu = {\rm det }(|\partial \bm{y}/\partial\bm{x}|)^{-1}$. Solving $\textrm{det}(|\partial \bm{y}/\partial \bm{x}|) = 0$ for a point source gives the image positions 
at which the magnification diverges, which collectively form the \textit{critical curves}. Mapping the critical curves to the source plane using the lens equation \eqref{eq:leqn} defines the \textit{caustic curves}. A point source on the caustic curve is expected to have infinite magnification. In practice, the maximum magnification is limited by the signal's wavelength and/or spatial size of the source. 
We refer the reader to App.~\ref{app:assump} for more details on the methods and assumptions used in the magnification calculation.

Evaluating the magnification as a superposition of point-like emitters covering the stellar surface, we find that the SMBHB lenses are able to magnify a star in the host galaxy of radius $10-10^3 R_{\odot}$ by a factor $\mu=10^{4-6}$ (Eq. \ref{eq:mag_ext_cusp_withunits} below). Bright sources, such as red giants or main-sequence (MS) O/B-type stars, have luminosities of $10^{2-6} L_{\odot}$ which can be amplified to $10^{6-12} L_{\odot}$, producing a potentially significant variability in the brightness of the entire host galaxy. In particular, given that massive O-type stars shine with the Eddington luminosity, $L_{\rm Edd}=4\pi G c M_*/\kappa=3.2\times 10^6L_{\odot} (M_*/100\Msun)$, where $\kappa$ is the electronscattering opacity, the luminosity variations of a lensed O-star matches $L_{\rm Edd}$ of a SMBH with $M_{\bullet}=\mu M_{*}=10^{7-8} \Msun (M_*/100\Msun)$. Thus, bright stars lensed by quiescent SMBHBs devoid of gas produce a bright active galactic nucleus (AGN), but with distinct spectral signatures and predictable variability patterns. 

The critical/caustic curves have three different topologies \cite{schneider:1986} based on $d\equiv |\bm{x}_{m_1}-\bm{x}_{m_2}|$, the dimensionless separation between the lenses projected into the lens-plane.
For eccentric binary lenses with semimajor axis $a$, eccentricity $e$, and orbital plane perpendicular to the line-of-sight to the observer, $d$ varies  between $(1+e)a/\xi_0$ (apocenter passage) and $(1-e)a/\xi_0$ (pericenter passage), where $a/{\xi_0} = 0.0075 \, T_{\rm yr}^{2/3}{M_{10}^{-1/6}D_{\rm kpc}^{-1/2}}$ from Kepler's law, $T_{\rm yr}$ is the orbital period in units of yr. For quasi-circular binary lenses with orbital plane inclination relative to the line-of-sight $\iota$, $d$ varies between $a/\xi_0$ and $a\cos{\iota}/\xi_0$.

\paragraph*{\bf \Qplsname.}

We consider a static source with respect to the lenses' center-of-mass, and focus on cases with a face-on 
binary lens with respect to the observer at separation $d\ll 1$. In this case, the caustic curve has a diamond shape with maximum diameter $\xi_0 d^2$ centered on the center-of-mass of the lenses, as shown in the Fig.~\ref{fig:master_diagram}. The corresponding critical curve is approximately a circle with radius $\approx \xi_0$ (the Einstein ring). Two additional caustic triangles form at distance $\sim \xi_0/d$ perpendicular to the binary axis see Fig.~\ref{fig:caustic} in App.~\ref{app:caus_size}. We will focus on the central diamond-shaped caustic, where the magnification is highest.

For a SMBHB on a circular orbit, the diamond-shaped caustic curve rotates rigidly around its center. 
The most prominent variability is obtained for a star situated within an annulus of inner and outer radii matching the innermost (fold) and outermost (cusp) points of the caustic, respectively. In this case, the caustic curve crosses the star multiple times during each period, resulting in high-magnification events 8 times (near-fold crossing) or 4 times (near-cusp crossing) per period. A significant quasi-periodic modulation is obtained even without caustic crossing 4 times per orbit in cases when the star is just outside the caustic. 
If the SMBHB inspirals due to GW emissions, the separation between the lenses shrinks, causing the caustic curve to shrink as $\sim d^2 \xi_0$ (see App.~\ref{app:caus_size}). Once the caustic is fully inside the star, the number of peaks and amplitude change in the light curve. 

For an equal-mass binary, the magnification for cusp and fold crossings are approximately between (see App.~\ref{app:mag} for more details)
\begin{align}
\label{eq:mag_ext_cusp_withunits}
 \mu_{\rm cusp} &\approx 1\times10^{6}(1\pm e)^{-0.7}M_{10}^{0.44}
 T_{\rm yr}^{-0.47} 
  R_{10}^{-0.64}D_{\rm kpc}^{0.67},\\
\label{eq:mag_ext_fold_withunits}
 \mu_{\rm fold} 
 &\approx 4\times10^{5}(1\pm e)^{-1}M_{10}^{5/12}
 T_{\rm yr}^{-2/3} 
 R_{10}^{-1/2}D_{\rm kpc}^{3/4},
\end{align}
where $e$ is the eccentricity, $R_{10}=R_{\rm src} /10 R_{\odot}$ is the radius of the source in units of 10 solar radii, $D_{\rm kpc} = D_{\rm LS}/{\rm 1\, kpc}$ is the lens-source distance in units of kpc, the $(1\pm e)$ dependence corresponds to caustic crossing near apastron and periastron passage respectively.
These estimates assume that the source diameter is smaller than the largest diameter of the caustic curve, i.e.
\begin{equation}\label{eq:Rsrc}
R_{\rm src}\leq \frac12 \xi_0d^2 \approx 3,500R_{\odot} \, (1\pm e)^2M_{10}^{1/6}     T_{\rm  yr}^{4/3} D_{\rm kpc}^{-1/2}    
\end{equation}
For larger sources, only a
portion of their area is highly magnified, resulting in a reduced overall magnification and a reduced level of variabiliy due to caustic rotation. 

The duration of peak brightening during caustic crossing is 
\begin{align}
    \label{eq:crossing_time}
    t_{\rm mag} \approx 16\, {\rm  hr} \,(1-e^2)^{-1/2} M_{10}^{-1/6} 
    T_{\rm  yr}^{-1/3} 
     R_{10} D_{\rm kpc}^{1/2}
\end{align}
for both apastron and periastron caustic crossing (see App.~\ref{app:crossing_time} for more details).

We will now present two examples for which \qpls{} by an inspiraling SMBHB produces prominent photometric variability: a quasi-circular SMBHB and an eccentric LISA binary. We will comment on the potential of these sources for multi-messenger observations.

\begin{figure*}[t]
    \includegraphics[width = 1\textwidth]{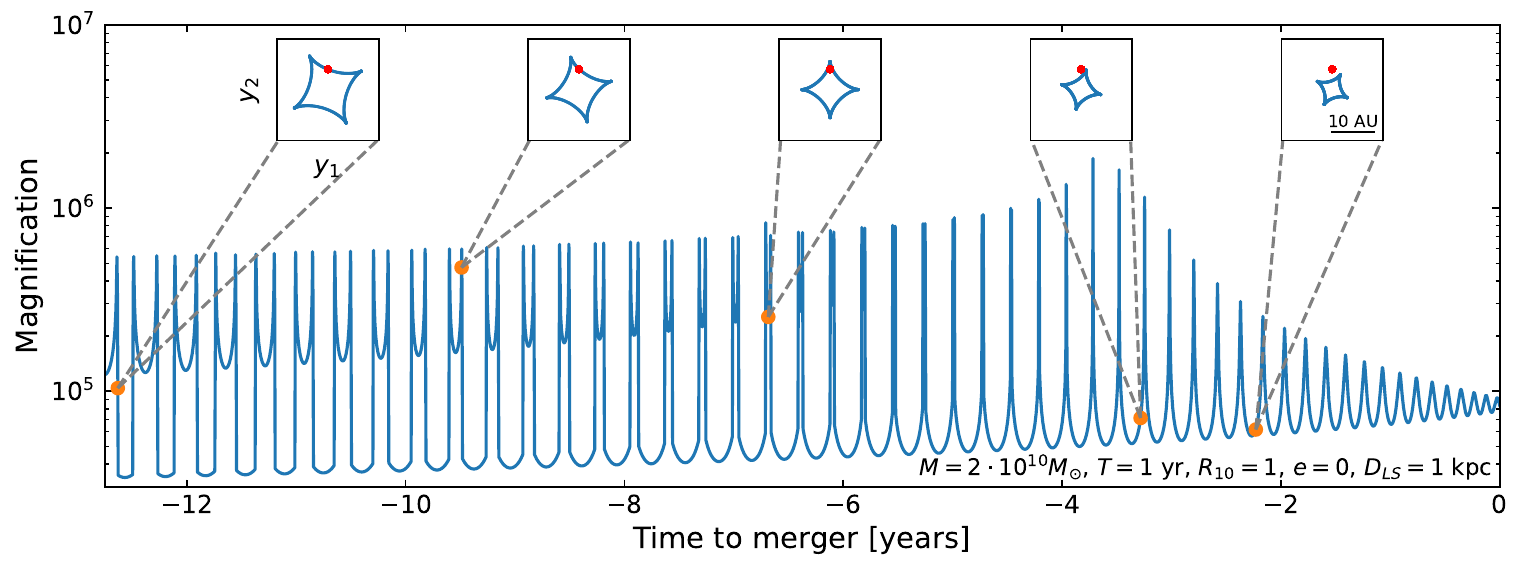}
    \caption{ 
    Magnification as a function of time-to-merger in \qpls{} by an inspiraling $10^{10} \Msun+10^{10} \Msun$ SMBH binary. The source star has a radius of $10 R_{\odot}$ and is 1 kpc from the binary, the initial binary period is 1 year producing a lensing spike every 2 months. The host galaxy is assumed to be at redshift $z=0.5$. The insets show the snapshots of the source plane with the fixed source star and the caustic curve, which rotates and shrinks during the inspiral. The source planes are plotted with the same scale shown in the last inset. Animations of the evolution of the SMBHB orbits, critical curves, images, caustic curves and the lightcurve are demonstrated in a GitHub repository \url{https://github.com/whanxi/SMBHB-Lensing-Animations}.
    \label{fig:massive_light_curve}}
\end{figure*}

\paragraph*{1) Quasi-circular merger \& pulsar timing arrays.}

Fig.~\ref{fig:massive_light_curve} shows the \qpls{} magnification curve of a star of radius $10 R_{\odot}$ in the host galaxy of an inspiraling quasi-circular, face-on, equal-mass SMBHB with $10^{10}\Msun$ component-mass binary lenses. 
The source star is fixed at a distance of 1kpc behind the SMBHB lens, 5 AU off-axis from the line-of-sight to the SMBHB center-of-mass in the source plane. The figure shows the flux magnification as a function of time to merger for the final 13 years of the inspiral before merger (the observed merger time is computed assuming the redshift of the host galaxy is $z=0.5$). All the observed times and frequencies of the \qpls{} sources are evaluated at this redshift in the remaining main text. Insets show 5 representative snapshots of the caustic curve which rotates counter-clockwise with the orbital period of the SMBHB. 

Magnification peaks occur when the caustic curve crosses the source, initially 8 times per orbit, i.e. once every 2 months at 13 yr to merger . Local minima occur when the source distance from the caustic has a maximum, i.e. when the source is aligned with the innermost points of the caustic (i.e. with the innermost fold, see Fig.\ref{fig:caustic} in App.~\ref{app:caus_size}). As the SMBH inspirals, the caustic curve shrinks in size (see insets). The light curve changes from fold crossing to cusp crossing producing the largest magnification $\mu_{\rm cusp}=1.8\cdot10^6$ at $3.8$ yr to merger. 
Afterwards, the source is entirely outside the caustic, producing 4 magnification peaks per orbit. The peak brightness decreases further as the binary approaches the merger.
However, the minimum brightness increases, approaching the magnification of the single source lens at merger, in this case $\mu=8\cdot 10^4$, see Eq.~\eqref{eq:mag_point_approx} in App.~\ref{app:minimum_mag}.

Gravitational waves from such massive binaries can be observed by pulsar timing arrays: an inspiraling $2\times 10^{10}\Msun$ system is coincidentally detectable to $z\lesssim 0.3$ by Nanograv 11yr and to very high redshift with future data~\cite[Fig.~4]{Kaiser:2020tlg}. 
Moreover, imposing a prior on the source position based on the \qpls{} detection increases the sensitivity with respect to all-sky GW searches~\cite{NANOGrav:2021sdv}. 

\begin{figure*}
    \includegraphics[width = 1\linewidth]{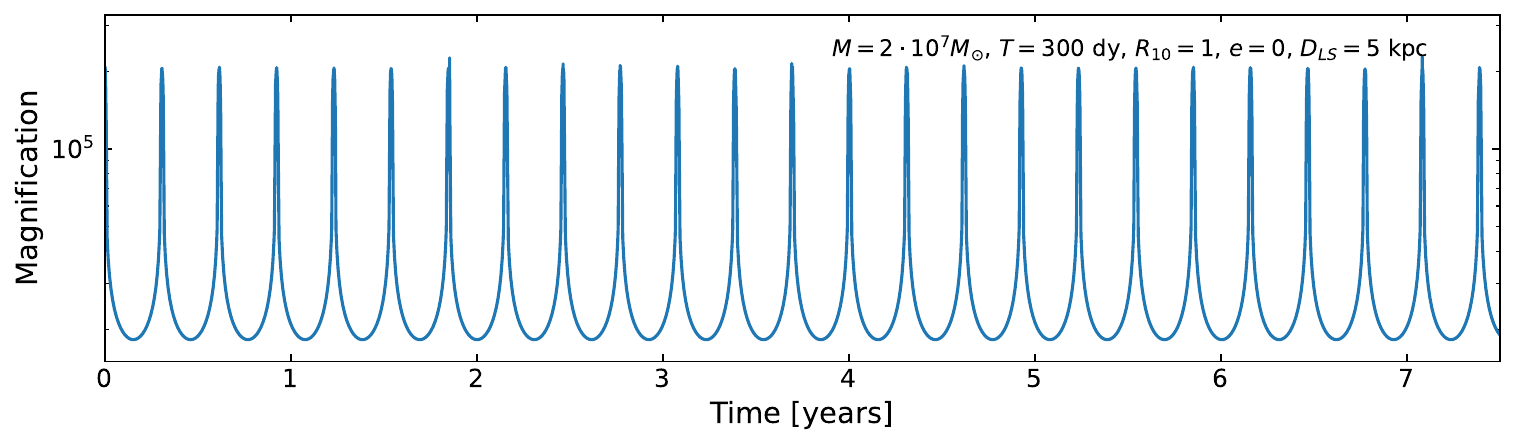}
    \includegraphics[width=1\linewidth]{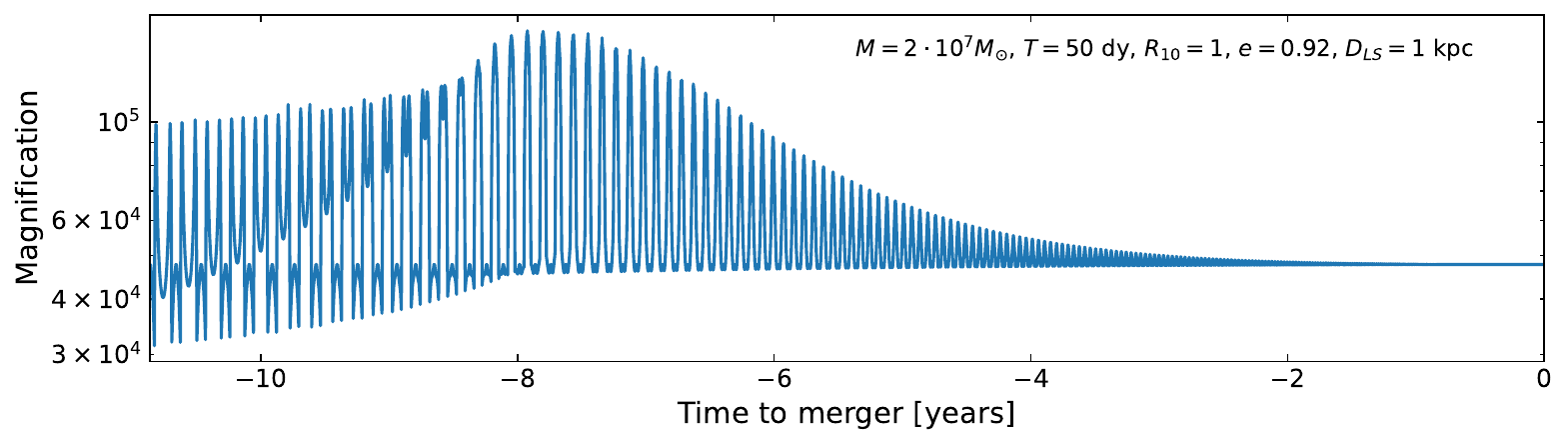}
        \includegraphics[width=1\linewidth]{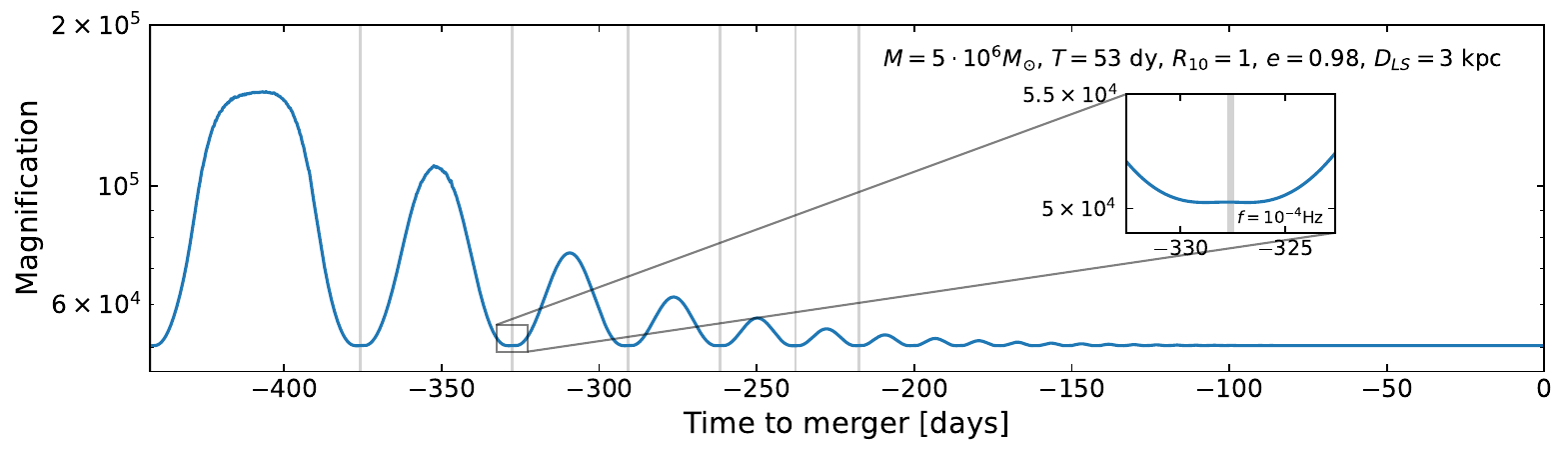}
    \caption{
    Lensing lightcurves by SMBH binaries in the LISA mass range. The host galaxies are assumed to be at redshift $z=0.5$. The total mass, initial period, initial eccentricity, source radius, source-lens distance are labeled in the plots, where $T_{\rm day} = T/{\rm day}$. \textit{Top}: a quasi-circular LISA binary $7.5\cdot10^5$ yr before merger representing the LISA binary population at an earlier stage. \textit{Middle}: an eccentric binary 11 years before merging in the LISA band. \textit{Bottom}: A highly eccentric LISA binary which enters the LISA band during each periastron passage represented by the grey lines. The inset shows one periastron passage, where the width of the gray line represents the few hours long time interval where the gravitational wave bursts are in the LISA band. \href{https://github.com/whanxi/SMBHB-Lensing-Animations}{\faGithub} }    \label{fig:lisa_light_curve}
\end{figure*}
\paragraph*{2) Eccentric binaries \& space-borne GW detectors.}
Let us now consider the synergy between high-magnification lightcurves and milli-Hz GW detectors, such as LISA and TianQin. These observatories are sensitive to systems with total mass $\lesssim 10^8 \Msun$. 

Fig.~\ref{fig:lisa_light_curve} shows three representative \qpls{} lightcurves for SMBHBs in the LISA/TianQin mass-range starting at $(0.75\,\mathrm{Myr}, 11\,\mathrm{yr}, 1.2\,\mathrm{yr})$ before merger with initial eccentricity $(0,0.92,0.98)$, respectively (see the legend for other parameters). These EM sources can inform the GW detection rate at a population level, provide an advance warning for future GW detections, and represent coincident multi-messenger observations in the three cases, respectively.
Caustic crossing produces a periodic magnification in the top panel. In the middle panel, the caustic curve initially crosses the source during apocenter passages. Due to the varying size of the caustic curve, there is only one instance of caustic crossing during each period. As the orbit and the eccentricity shrink, the lightcurve initially exibits doubly peaked fold crossing, which then changes to a single-peaked cusp crossing and the modulation amplitude decays as the caustic shrinks and the source is fully outside of the caustic.  For this system, the SMBHB enters the LISA band a few years after prominent \qpls{} modulation. The GW frequency peaks at
$0.1$ ($0.02$) mHz 0.8 days (5.5 months) before merger at which point the eccentricity is down to 0.05 (0.45). The bottom panel shows the case of a highly eccentric $e=0.98$ SMBHB lens being a LISA GW-source, where \qpls{} magnificantion peaks are initially again near the apocenter. 
In this case, the GW bursts emitted during pericenter passages (troughs of the EM lensing light curve) are already in the detectable frequency band (see inset and gray vertical lines) one year before merger.
The LISA signal-to-noise ratio (SNR) for a single burst at pericenter reaches 10 at a cosmological redshift of $z\approx0.5$ \cite{Xuan2024}. The LISA SNR accumulates mainly in the last months before merger; it reaches 10 at $z\approx4.9$ (see App.~\ref{app:time_to_merge} for more details).

\begin{figure}
    \centering
    \includegraphics[width=1\linewidth]{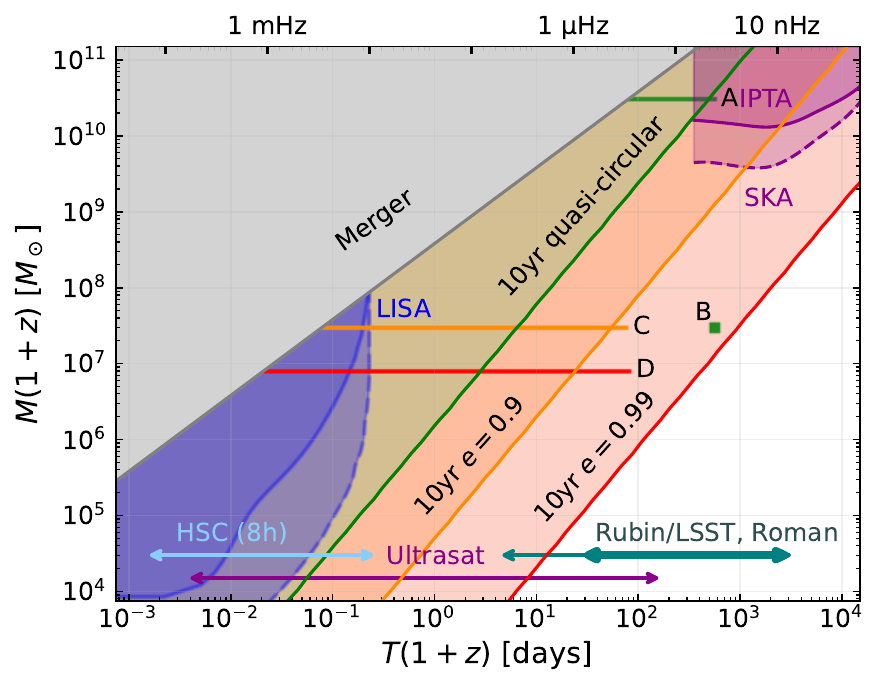}
    \caption{Prospects for multi-messenger observations of SMBHBs using \qpls{}  in the plane of total SMBH mass and redshifted orbital period. Slanted filled regions show the systems that will merge in 10 years for different values of the eccentricity. Horizontal lines correspond to the systems shown in Figs.~\ref{fig:massive_light_curve} (A) and \ref{fig:lisa_light_curve} (B,C,D - top to bottom) at $z=0.5$.
    LISA regions show the orbital period where the signal-to-noise-ratio above $10^{-4}$ Hz reaches $>10$ for equal-mass, non-spinning and quasi-circular SMBHBs at $z=0.5$ (dashed) and $z=5$ (solid). 
    PTA regions correspond to IPTA 2025 (solid) and SKA 2034 (dashed) for quasi-circular q=1 SMBHs at z=0.5~\cite{Charisi:2021dwc}. 
    Horizontal arrows show the cadence-to-survey time of several transient surveys (these apply for all masses). \change{For Rubin/LSST we distinguish the cadence of any-filter (thin, 4 days) from that of all-6-filters (thick, 24 days). 
    A circular (eccentric) binary produces 4 (1) single-peaked or double-peaked caustic crossings during each orbital period. 
    } 
    }
    \label{fig:Lisa_Lsst}
\end{figure}
Fig.~\ref{fig:Lisa_Lsst} summarizes the parameter space for multi-messenger \qpls{} observations in terms of the SMBHB total mass and orbital period. 
Green, orange, red lines correspond to sources merging in 10 years for different values of the initial eccentricity (0, 0.9, 0.99), respectively. Vertical/horizontal lines indicate the range of GW/EM facilities. The source presented in Fig.~\ref{fig:massive_light_curve} is indicated with (A), and those in Fig.~\ref{fig:lisa_light_curve} are marked with (B), (C), and (D), respectively. {In many cases, caustic crossing of stars by SMBHBs} observable by mHz GW detectors probe the same underlying population, but decades to millions of years before merger.
Some of the most massive quasi-circular LISA sources $(1+z)M\gtrsim 10^{7.5} \Msun$ could represent delayed multi-messenger events, where the lensing lightcurve arrives only a few years in advance, which can be detected by surveys such as Rubin-LSST which has a cadence of 
\change{$\sim 4$} days (for a single filter) \cite{LSST:2008ijt}.
Indeed, the observed time to merger as a function of observed period for circular sources is $t_{\rm mg,c}=8\,\mathrm{yr} (T_{20\rm day})^{-8/3} M_{8z}^{-5/3} [4q/(1+q)^2]^{-1}$ where $q$ and $M_{8z}=(1+z)M/(2\times 10^8\Msun)$ represent the mass ratio and the redshifted total mass of the SMBHB, respectively \cite{Peters1964}.

For quasi-circular SMBHs, coincident detections of \qpls{} with mHz GWs requires not only high-cadence EM surveys (e.g. 30 min for Subaru HSC-SSP \cite{HSC2018b}), but also sub-solar source diameters. Indeed, GW frequencies of $f_{\rm GW}\gtrsim 0.1\,$mHz correspond to binary orbital periods of $T(1+z)=1/(2f_{\rm GW}) \lesssim 4\,$hr, the size of the caustic curve is smaller than $0.1 R_{\odot} M_{10}^{1/6}D_{\rm kpc}^{-1/2}$, and the magnification is suppressed for intrinsically larger $R_{\rm src}$ EM sources or for sources with smaller $D_{\rm LS}$ (see Eqs.~\eqref{eq:mag_ext_cusp_withunits}--\eqref{eq:Rsrc}).
Detectable \qpls{} sources may be limited to the nuclear star cluster of the SMBHB lens for main sequence stars or else include the lensing of compact objects, such as white dwarfs, radio pulsars, or X-ray binaries.
Bright inhomogeneities of larger sources (e.g. starspots or flares of O/B stars) may also be significantly magnified for circular SMBHBs emitting in the LISA band.

Coincident EM and GW transients are easier to achieve for highly eccentric SMBHBs, as the merger time scales approximately as $t_{\rm mg, e}=(1-e^2)^{7/2} t_{\rm mg, c}$. This allows systems with total mass $\lesssim 10^7\Msun$ to evolve from $T\sim$ days (LSST cadence) to merger in a decade or less. There are multiple reasons to potentially expect eccentric SMBHBs.
The scattering of stars in the vicinity of the SMBHB \cite{Aarseth:2008,Sesana:2008,Berentzen:2009,Iwasawa:2011,Bonetti+2020,Khan+2021,Mastrobuono-Battisti+2025,Gualandris+2022} and/or the interaction with a gaseous disk \cite{Goldreich_Sari2003,Cuadra:2008xn,Roedig+2011,Siwek+2024} if present promotes eccentricity growth. Cosmological simulations followed by N-body simulations \cite{Chen:2024,Zhou:2025,Mukherjee:2025} also show evidence of a large number of merging massive binary systems with high eccentricity.
The orbital eccentricity also leads to observable imprints on the \qpls{} lightcurve, as shown in the center and lower panels in Fig.~\ref{fig:lisa_light_curve}.
The binary separation becomes larger at apocenter passage, allowing for relatively larger stars to be highly magnified, compared to the quasi-circular SMBHBs.
The slow velocity at apocenter passage also increases the duration of each peak in the lightcurve. 
These factors make a coincident or delayed multi-messenger observation more promising for an eccentric SMBHB. Similar effects of coincidental quasi-periodic GW and EM bursts are found in general relativistic magnetohydrodynamic (GRMHD) simulations of eccentric SMBHBs with gas accretion \cite{Manikantan2025}. This effect can be distinguished from \qpls{} by analysing the spectrum.

\paragraph*{\bf Conclusions and Prospects.}

We have described \qplsname{} (\qpls): how a slowly moving bright star lensed by a SMBHB imprints a modulation of the light-curve at the center of the host galaxy, offering a direct probe of the binary's orbit and its evolution. 
The lensing of individual bright stars result in spectacularly bright EM sources comparable to that of AGN even if the SMBHB lens resides in a quiescent galactic nucleus, in some cases outshining the light of the entire host galaxy. \change{The \qpls{} light curve is a result of a combination of the binary orbit and the source’s motion. Valuable information is contained in the separation between peaks, and width/amplitude of each peak, which is absent in many other transients. The spectrum (e.g. emission/absorption lines) of the source can also be used to distinguish lensed stars from other transients.\footnote{\change{A spectral followup of the following peak(s) could be useful to identify a change in the spectral signature during the caustic crossing to reveal prominent stellar emission/absorption lines of the lensed starlight that were weaker before/after.}}} Very massive or eccentric SMBHBs can be also studied via GWs (PTAs or space detectors), either concurrently with \qpls{} or a few months/years after if the EM lightcurve provides an advance warning of the merger.
This type of observation will enable ``classical'' multi-messenger applications, including host identification, cosmography~\cite{Schutz:1986gp} and tests of general relativity~\cite{LISACosmologyWorkingGroup:2019mwx}. 
\qpls{} lightcurves may also reveal tight SMBHBs also at an earlier evolutionary stage even without any GW detections.

Rapid developments in transient astronomy provide abundant opportunities to search for SMBHBs in this way: 
Zwicky Transient Facility observes variability over scales of minutes-years~\cite{Graham:2019qsw}, 
Subaru's Hyper-Supreme Cam can preform targeted searches with cadence of minutes~\cite{HSC2018a,HSC2018b}.
The Rubin Observatory will monitor $\sim 2\times 10^{10}$ galaxies over ten years~\cite{LSST:2008ijt,LSSTScience:2009jmu}, with time resolution $\sim$ days (single filter). The Roman Space Telescope will search the sky with a smaller survey area but higher resolution and deeper imaging~\cite{roman_overview}.
It is able to observe periodic signals of SMBHB with a cadence as low as 5 days to a depth of $\sim26$ mag \cite{romanMBHB:2023}.  ULTRASAT will monitor ultraviolet variable events on scale of minutes to months~\cite{Shvartzvald:2023ofi}. In addition to wide-field seaches, further sensitivity to SMBHs can be achieved by dedicated observations of candidate systems. 

The probability of a \qpls{} in a SMBHB can be estimated as the integral of the source density over the cylinder encompassing the central caustic, weighted by the probability of the source to remain within the caustic for at least an orbital period.
Assuming a Gaussian velocity distribution and choosing a minimum $D_{\rm LS}$ such that $\xi_0\geq a$ gives $P_{\rm QPLS} \sim 1.5 \cdot 10^{-5} \left(\frac{n_\star}{{\rm pc}^{-3}}\right) T_{\rm yr}^{8/3} M_{10}^{1/3} F_Q$, where the stellar density $n_\star$ is considered homogeneous and $F_Q$ is a logarithmic correction that depends on $M,T$ (see App.~\ref{app:qpls_prob}). 
Extrapolating these rates to synthetic SMBHB catalogs~\cite{Barausse:2023yrx} gives $1-50\; [190-5,000] \left(\frac{n_\star}{{\rm pc}^{-3}}\right)$ \qpls{} sources in systems with a period below $10 \; [40]$ yr (observer frame) at $z<0.3$. These rates consider only comparable-mass systems $M_2/M>0.5$, a rather small fraction of the total.

The presence of SMBHBs can be inferred from stars crossing the binary's caustic network, even for binaries with long period $T\gg 10$ yr or with stars that are too fast for multiple crossings due to orbit.
For an individual SMBHB, the caustic crossing rate is $1.9\cdot 10^{-4}  {\rm yr}^{-1} M_{10}^{1/3} T_{\rm yr}^{5/3}\left(\frac{n_\star}{{\rm pc}^3}\right) F_1$, where $F_1(\tilde D, M, T)$ is a logarithmic correction.
(see App.~\ref{app:caustic_crossing_rate} for more details, including crossings of secondary caustics). 
Extending this estimate to synthetic SMBHB populations~\cite{Barausse:2023yrx}, we find a total rate of $300-10^5 \left(\frac{n_\star}{{\rm pc}^{-3}}\right) \, [{\rm yr}^{-1}]$  
caustic crossings per year in all galaxies within $z<0.3$. The rates includes only comparable-mass SMBHBs $M_2/M>0.5$ at separations less than $10^3 GM/c^2$; the rates increase greatly when including wider separations, higher redshifts or more asymmetric binary masses (see App.~\ref{sec:probability_catalog}).  
Considering only SMBHB populations in agreement with the high-$z$ quasar luminosity function leads to rates $\geq 10^4 \left(\frac{n_\star}{{\rm pc}^{-3}}\right) \mathrm{yr}^{-1}$.
These estimates warrant a more detailed analysis of the impact of SMBHB astrophysics, stellar population, and the sensitivity of time-domain surveys.

Realistic \qpls{} lightcurves require including additional effects, such as combining the peculiar motion of the star with the orbit of the SMBHB. Observing the lensing lightcurve will be easiest if the SMBHB lens is not embedded in gas. Gas around the SMBHB may lead to additional signatures: opaque regions may block some but possibly not all of the source's images as they move around the critical curve, potentially providing information on the size and structure of accretion disks. Indeed, there are typically two to four highly magnified images separated by $1.4\mathrm{pc}M_{10}^{1/2}D_{\rm kpc}^{1/2}$ (Eq.~\ref{eq:ein_rad}). \qpls{} is also sensitive to other properties of SMBHBs, such as the mass ratio, and may be able to indicate the presence of a third SMBH. The tendency of luminous stars to form clusters provides additional challenges and opportunities: i.e. multiple sources may need to be modeled but they may also probe the spatial structure of the lens. These effects may help distinguish SMBHBs lensing imprints from other processes that induce variability in a galaxy's lightcurve.

The high magnification and persistence of \qpls{} modulations may allow even moderately luminous stars to reveal SMBHBs. 
Even a Sun-like star ($1L_\odot, 1R_\odot$) magnified by the SMBHB system similar to that in Fig.~\ref{fig:massive_light_curve} but with arbitrary mass may be detectable by Rubin-LSST (Roman) at a distance $\lesssim 80 \; (400) \,M_{10}^{5/24}T_{\rm yr}^{-1/3}(R_{10}/0.1)^{-1/4} {(L/L_{\odot})^{1/2}}D_{\rm kpc}^{3/8} \, \text{Mpc}$
(based on single exposure sensitivity of 24 (27.63) mag at 750nm and Eq.~\ref{eq:Dl_horizon}, cf.~App.~\ref{sec:detection_thresholds}). 
Photometric errors scale with the number of visits as $\propto 0.0082 \sigma_{m, \rm single}\sqrt{1{\rm yr}/t_{\rm obs}}$,  and can be further improved by knowing the lightcurve: matched filter techniques, similar to searches for GWs, may be used to recover \qpls{} signals from noise-dominated data. Additional gains may follow from spatial information (e.g.~from pixel-level microlensing~\cite{1992_Crotts_pixel_microlensing_idea,2001Lewis_pixel_microlensing_extragalactic,Gil-Merino:2006ypx}) and spectral features, in addition to photometry, and from targeting galaxies more likely to harbor SMBHBs with very large telescopes. 
These and other advances in observations and data analysis will allow \qpls{} to open a new window to illuminate the most extreme objects in the heart of distant galaxies.

\begin{acknowledgments}
We acknowledge valuable discussions with Bruce Allen, Jonathan Gair, Juan Garcia-Bellido, Graham Smith and Stephen Taylor.
This work was supported by the Science and Technology Facilities Council Grant Number ST/W000903/1 (to BK). 
\end{acknowledgments}

\appendix 

\section{Methods and Assumptions}\label{app:assump}

In this section, we explain the basic methods and assumptions we use to compute the magnification of a star lensed by the SMBHB. We justify that we can treat the lenses as two point-mass lenses on a single plane without considering the microlensing effect from other stars in the same galaxy and the macrolensing effect of the galaxy.

\subsection{Weak field approximation}

In this work, we use the weak-field lensing approximation. This is justified because $\Phi/c^2 \sim R_g/\xi_0 \sim 3\cdot10^{-4} \sqrt{M_{10}/D_{\rm kpc}}$ is small for our system. Here $\Phi$ is the gravitational potential, evaluated on the main critical curve (Einstein radius). 
Strong field corrections could become important for sources near the secondary (triangle) caustic, as images form much closer to the binary lens, which can be used as a test of gravity \cite{Zhong2024}. We will defer this matter to a future study.

\subsection{Point-mass lenses}

In this work, we only consider the lensing effect from the SMBHB and ignore the lensing effect of the host galaxy.
If we model the galaxy with an isothermal singular sphere, the Einstein radius is given by \cite{Schneider:1992}
\begin{equation}
    \label{eq:ein_rad_galaxy}
    \xi_{\rm galaxy} = \frac{4 \pi \sigma^2D{_L}D_{\rm LS}}{c^2 D_{S}}.
\end{equation}
Using Eq.~\eqref{eq:ein_rad}, and assuming a $M-\sigma$ relation of 
\cite{Kormendy2013} 
\begin{equation}    \label{eq:M_sigma}    M = 3.1\times 10^8 \Msun\left(\frac{\sigma_v}{200 \,{\rm km/s}}\right)^{4.38},
\end{equation}
where $\sigma_v$ is the velocity dispersion of the stars. The ratio of Einstein radius of the SMBHB (Eq.~\eqref{eq:ein_rad}) to that of the galaxy is given by 
\begin{align}
    \label{eq:ein_rad_ratio}
    \frac{\xi_{\rm BH}}{\xi_{\rm galaxy}} =\frac{c}{2 \pi \sigma^2}    \sqrt{\frac{GM}{D_{\rm LS}}}
    =
    51\,M_{10}^{0.04} D_{\rm kpc}^{-1/2}.
\end{align}
The SMBHB lens dominates over the galaxy lens because of the short lens-source distance as compared to a source and a lens in different galaxies; the dependence on the mass of the SMBHB is weak. In the \qpls{} examples presented in the main text, we have the source positioned near the caustic curve, with the source position in units of the Einstein radius of the SMBHB lens approximately $y_{\rm BH} \sim 10^{-5}$. The corresponding magnification caused by the SMBHB lens is approximately $\mu_{\rm BH} \sim 10^5-10^6$. From Eq.~\eqref{eq:ein_rad_ratio}, if we consider the host galaxy as the lens instead, the source position in units of the Einstein radius of the galaxy is approximately $y_{\rm galaxy} \sim2\cdot10^{-3}$. The point-source magnification of the galaxy lens is given by $\mu_{\rm galaxy} \sim 1/y_{\rm galaxy}\sim 10^{3}$, much less than that caused by the SMBHB lens. We thus ignore the contribution from the galaxy lens when evaluating the light curves.

The additional lensing effects of stellar-mass objects can be neglected when calculating the lensing by SMBHBs. The reason is that sources in the same galaxy as the SMBHB lens lead to very low convergence for the stellar-mass objects
\begin{equation}
\kappa_* \equiv \frac{\Sigma_{*}}{\Sigma_{\rm cr}} = 6\cdot 10^{-5} \frac{\Sigma_*}{10^5 \Msun/{\rm pc}^2} D_{\rm kpc}\,,
\end{equation}

where $\Sigma_{\rm cr}=\frac{D_S}{4\pi G D_L D_{\rm LS}}=1.66\cdot 10^9 \Msun/{\rm pc}^2\,D_{\rm kpc}^{-1}$ is the critical surface density for lensing \cite{Schneider:1992} and $\Sigma_*$ is surface density of stars in the galaxy at distance $D_{\rm LS}$ from the center. $\kappa_*$ is very small even for the given reference value, which is very close to an empirical upper limit observed across many systems~\cite{maximum_surf_density}. This situation is in stark contrast to galaxy/cluster strong lenses, where $D_L, D_S, D_{\rm LS}\sim \mathcal{O}(\text{Gpc})$ reduce $\Sigma_{\rm cr}$ by a factor $\sim 10^6$, widening the caustic curves and reducing the maximum magnification by several orders of magnitude, even for modest $\Sigma_*\sim 20 \Msun/{\rm pc}^2$~\cite{Diego:2017drh,Oguri:2017ock,Venumadhav:2017pps,Muller:2024pwn}.

\subsection{Single-plane lenses}

In this work, we project both lenses to a single lens plane instead of evaluating the lens equations with two lens planes where the two point lenses sit. Following the double-plane prescription of Ref.~\cite{Schneider:1993}, we label the lens planes of the closer and farther SMBHs to the observer, respectively, as the lens-planes-1 and -2, and set the origin of lens-plane-1 to the position of the closer SMBH and define $\bm{x}^{(2)}_{m_2}$ to be the position of the farther SMBH in lens-plane-2. The two lens planes are parallel, and they are perpedicular to the line-of-sight to the observer. The double-plane lens equations are given by \cite{Schneider:1993}
\begin{align}
    \bm{y}& = \bm{x} - m_1 \frac{\bm{x}}{|\bm{x}|^2} -m_2 \frac{\bm{x}^{(2)}-\bm{x}^{(2)}_{m_2}}{|\bm{x}^{(2)}-\bm{x}^{(2)}_{m_2}|^2} \, , \label{eq:double_plane_lens_eq1} \\
    \bm{x}^{(2)} &= \bm{x} - m_1 \beta \frac{\bm{x}}{|\bm{x}|^2}\, ,\label{eq:double_plane_lens_eq2}
\end{align}
where we define
\begin{align}
    m_1 = \frac{M_1}{M_1+\frac{D_{2S}D_L}{D_{\rm LS}D_{2}}M_2} \, , \quad
       m_2 &= \frac{M_2\frac{D_{2S}D_L}{D_{\rm LS}D_{2}}}{M_1+\frac{D_{2S}D_L}{D_{\rm LS}D_{2}}M_2}\, ,\label{eq:m12_def} 
\end{align}
$\beta = {D_{12}D_S}/{D_{\rm LS}D_2}\approx {D_{12}}/{D_{\rm LS}}$ for the source and lenses in the same galaxy, 
$D_{12}$, $D_{2S}$, $D_2$, and $D_S$ are respectively the distance between the two lens planes, between the source and lens-plane-2, between lens-plane-2 and the observer, and between lense-plane-1 and the observer, and $D_{\rm LS}=D_{2S}+D_{12}$, where we assume $D_{12}\ll D_{2S}\sim D_{\rm LS} \ll D_2 \sim D_{S}$. A light ray emitted at the source position $\bm{y}$ 
crosses
lens-plane-2 at $\bm{x}^{(2)}$ and then crosses lens-plane-1 at $\bm{x}$, which 
is the image position as seen by the observer.\footnote{Note that $\bm{y}$ and $\bm{x}$ are 2D vectors describing the direction to the physical source position and its apparent image, respectively, in the planes perpendicular to the line-of-sight to the observer.} Here $\beta \leq a/D_{\rm LS}\approx 10^{-5} M_{10}^{1/3}T_{\rm yr}^{2/3}D_{\rm kpc}^{-1}\ll 1$ 
since $D_{12}\leq a \ll D_{\rm LS}$, $a$ being the SMBHB separation. 
Hence the second lens equation Eq.~\eqref{eq:double_plane_lens_eq2} is approximately $\bm{x}^{(2)}\approx\bm{x}$, and the light ray crosses both lens planes at approximately the same position. The first lens equation in Eq.~\eqref{eq:double_plane_lens_eq1} is thus approximately equivalent to the single-plane lens equation in Eq.~\eqref{eq:leqn} in the main text. The SMBHB lensing system may be approximated with having a single lens plane because of the close proximity of the two SMBH components and because the source and SMBHB are in the same host galaxy far from the observer.

\subsection{Lens equations and magnification}

As argued above, we can model the SMBHB as two point masses in the same plane. The lens equation of a general lens is given by 
\begin{equation}
    \label{eq:lens_eq_general}
    \bm{y} = \bm{x} - \bm{\nabla}\psi(\bm{x}),
\end{equation}
where $\psi(\bm{x})$ is the dimensionless gravitational lensing potential by projecting the three-dimensional Newtonian potential to the lens plane \cite{maneghetti:2022book}. In particular, for a binary lens, 
\begin{equation}\label{eq:psi}
\psi(\bm{x}) = m_1\log|\bm{x}-\bm{x}_{m_1}|+m_2\log|\bm{x}-\bm{x}_{m_2}|    
\end{equation}

Substituting this potential in Eq.~\eqref{eq:lens_eq_general} gives the lens equation of a binary lens in Eq.~\eqref{eq:leqn} in the main text. The magnification of a point source is given by ${\rm det} (\mathbf{A})^{-1}$, where
\begin{equation}
    \label{eq:A_def}
    \mathbf{A} = \frac{\partial \bm{y}}{\partial \bm{x}},
\end{equation}
which is calculated using Ref~\cite{Bozza2010,Bozza2021} \footnote{\url{https://github.com/valboz/VBMicrolensing}}. 

\subsection{Magnification of extended sources and time delay}

We evaluate the magnification of an extended source by evaluating the point-source magnification in a $200\times200$ grid covering the cross section of the source in the source plane. We compute the average point-source magnification of the grid and sum up the contribution from all images. 
This is justified because the time delay between the highly magnified images during a caustic crossing is small.

 When a source is close to the caustic curve on the interior side of it, the highly magnified images are close to the critical curves and close to each other. Consider a source near a fold, for example, the two most magnified images near the critical curve are located at $\Delta x=\pm \sqrt{\Delta y/3\tilde{h}}$
 \cite{Schneider:1992}, where we transform the coordinates such that the $x$ and $y$ axes are aligned with the local critical or caustic curve. $\Delta x$ and $\Delta y$ measure the perpendicular distance of the image and source from the nearby critical or caustic curve. 
 See App.~\ref{app:mag} for more details of the coordinate transformation and the definition of $\tilde{h}=6\,\partial^3_{x_2} \psi(\bm{x})$ evaluated at the critical curve. The time delay of a binary lens is given by
\begin{align}
T_d(\bm{y},\bm{x}) = &\frac{D_S \xi_0^{^2}}{cD_LD_{\rm LS}} \left[\frac 12 |\bm{x}-\bm{y}|^2 - \psi(\bm{x})\right]\label{eq:delay}
\end{align}

For tight binaries with $d\equiv |\bm{x}_{m_1}-\bm{x}_{m_2}|\ll1$, the critical curve is approximately a circle of radius 1, with $\tilde{h}\approx d^2/8$ (see Eq.~\eqref{eq:hbar_approx_fold}). Hence, the two images are at $x\approx1\pm \sqrt{8y/3d^2}$. For \qpls{} events with large magnification modulation, the source is near the annulus bounded by the cusps ($y=\frac12 d^2$) and the innermost fold points ($y=\frac14 d^2$), thus $y\sim d^2$ applies. The lens positions are at $\bm{x}_{m_{1,2}} =\pm d/2$ for equal lens masses. For small $d$, we can approximate both the source position and the lens position to be at 0. The three distance terms in parentheses of Eq.~\eqref{eq:delay} are approximately $1\pm \Delta x$. Expanding to first order in $\Delta x$ and computing the difference between the two image positions, the terms inside the bracket are thus of order $\Delta x\sim\sqrt{\Delta y}/d$. 
During a caustic crossing, the largest distance (largest time delay) between the extended source and the caustic curve is set by the size of the source $\Delta y\sim R_{\rm src}/\xi_0$. 
This leads to a characteristic time delay between the two images of order
\begin{align}
    \label{eq:delay_approx}
    \delta T_d &\sim \frac{\xi_0^{3/2}R_{\rm src}^{1/2}}{cD_{\rm LS}d} = \frac{\xi_0^{5/2}R^{1/2}}{cD_{\rm LS}a}
    \notag\\
    &\sim3\, {\rm hr} M_{10}^{13/12} 
    T_{\rm  yr}^{-1/3} 
     R_{10}^{1/2} D_{\rm kpc}^{-1/4},
\end{align}
which is small compared to the primary caustic crossing time computed in Eq.~\eqref{eq:crossing_time} in most of the systems that we considered. 

Both highly magnified images are prominent and appear with negligible time difference during each caustic crossing at the fold. We also find a small time delay for all 3 highly magnified images during caustic crossing at the cusp. Even if one image is not detected, the magnification is still of the same order as estimated because all 2/3 highly magnified images are of similar magnification. As long as the time delay between images is much shorter than the orbital period of the SMBHBs, we do not lose the quasi-periodic feature of the light curves.

Note that images other than the 2/3 most magnified ones might have very different impact parameters, approximately $\sim \xi_0$, see Fig.~\ref{fig:master_diagram}. In tight binaries, the 2/1 less magnified images are still close to the Einstein ring and have a high magnification. In Fig.~\ref{fig:master_diagram} for example, the three top images are the most magnified induced by the nearby cusp. The lower image is less magnified, but still relatively close to the Einstein ring. In this case, the typical time delay of the subdominant images is approximately $\sim GM/c^3\sim14 \,{\rm hr} \cdot M_{10}$, which is longer than the lensing peak duration but still much shorter than the orbital period. We find that the magnification of the subdominant images is $\sim10^{-2}-10^{-1}$ of the peak magnification during caustic crossing. Hence, the longer time delays of the subdominant images will not destroy the repeating lensing peaks. 

This geometry also shows that if the SMBHB is surrounded by obscuring gaseous clouds or a torus of characteristic size less than $\xi_{0} = 1.38\text{pc}\times M_{10}^{1/2}{D_{\rm kpc}^{1/2}}$, the occultation of some of the images may or may not lead to a significant reduction of the overall light depending on whether or not all 2/3 dominant images are occulted.

\section{Caustic curve topology and size}\label{app:caus_size}
We find the critical curve and the caustic curve by evaluating ${\rm det} (\mathbf{A}) = 0$ following Ref.~\cite{maneghetti:2022book}. Expressing the source position with a single complex number using the Cartesian components of the image position, $z = x_1+ i x_2 $, the equation ${\rm det} (\mathbf{A}) = 0$ may be expressed as the norm of a function of $z$ is equal to one. Setting this to $e^{i \theta}$ where $\theta$ is an arbitrary parameter, the critical points for $m_1=m_2$
are given by solving the equation
\begin{equation}
    \label{eq:crit_eqn}
  z^4-\left(\frac{d^2}{2}+e^{i \theta}\right)z^2-\frac{d^2}{16}(d^2-4e^{i\theta})=0,
\end{equation}

Solving for all possible solutions for $z$ (up to 4 solutions for each value of $\theta$) in the entire range of values of $0\leq\theta<2\pi$ gives all the critical points. For the general equation of the critical curve with $m_1\neq m_2$, see Ref.~\cite[Eq.~(4.98)]{maneghetti:2022book} for more details. The critical and caustic curves have three different topologies depending on the separation $d$ and the mass ratio $q =m_1/m_2$.

\begin{figure}
    \centering
    \includegraphics[width=1\linewidth]{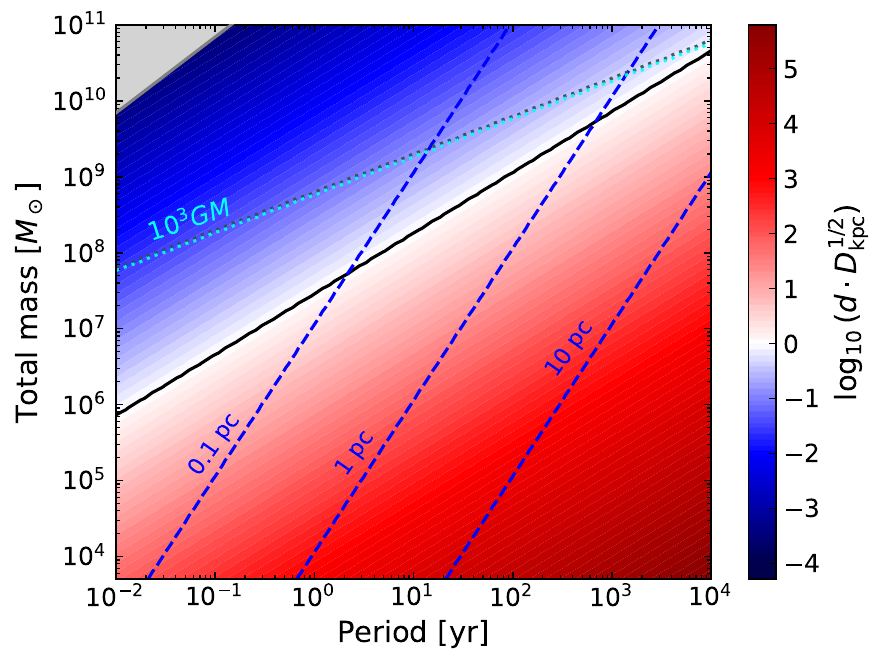}
    \caption{Dimensionless orbital separation $d=a/\xi_0$ (Eq.~\ref{eq:d_dimensionless_scale}) of a binary lens if the source is at 1\,kpc. Dashed lines represent constant physical orbital separation. The dashed line indicates the transition to GW-dominated inspiral at $a=10^3 GM/c^2$~\cite{Haiman_Kocsis_Menou2009}.}
    \label{fig:caustic_size}
\end{figure}

The most important parameter characterizing the properties of lensing by a binary in Eq.~\eqref{eq:crit_eqn} is the dimensionnless binary separation in units of Einstein-radii
\begin{equation}\label{eq:d_dimensionless_scale}
    d \equiv \frac{a}{\xi_0} \approx
    \frac{a}{2\sqrt{R_{\rm g} D_{\rm LS}}} = 0.0075 \, \frac{T_{\rm yr}^{2/3}}{M_{10}^{1/6}D_{\rm kpc}^{1/2}}\,.
\end{equation}
Here $R_{\rm g}=GM/c^2$ and Kepler's law has been assumed to relate the semimajor axis and binary's period, $a = 0.01
\, \text{pc} \; M_{10}
^{1/3}T_{\rm yr}^{{2}/{3}}$.
Figure \ref{fig:caustic_size} shows how $d$ varies for the different masses of the SMBHB lens and orbital period. SMBHBs observable on transient survey timescales, pulsar timing array and LISA are typically in the tight-binary regime. For this reason, we will focus on the $d\ll 1$ limit.

\begin{figure*}
    \includegraphics[width=1\linewidth]{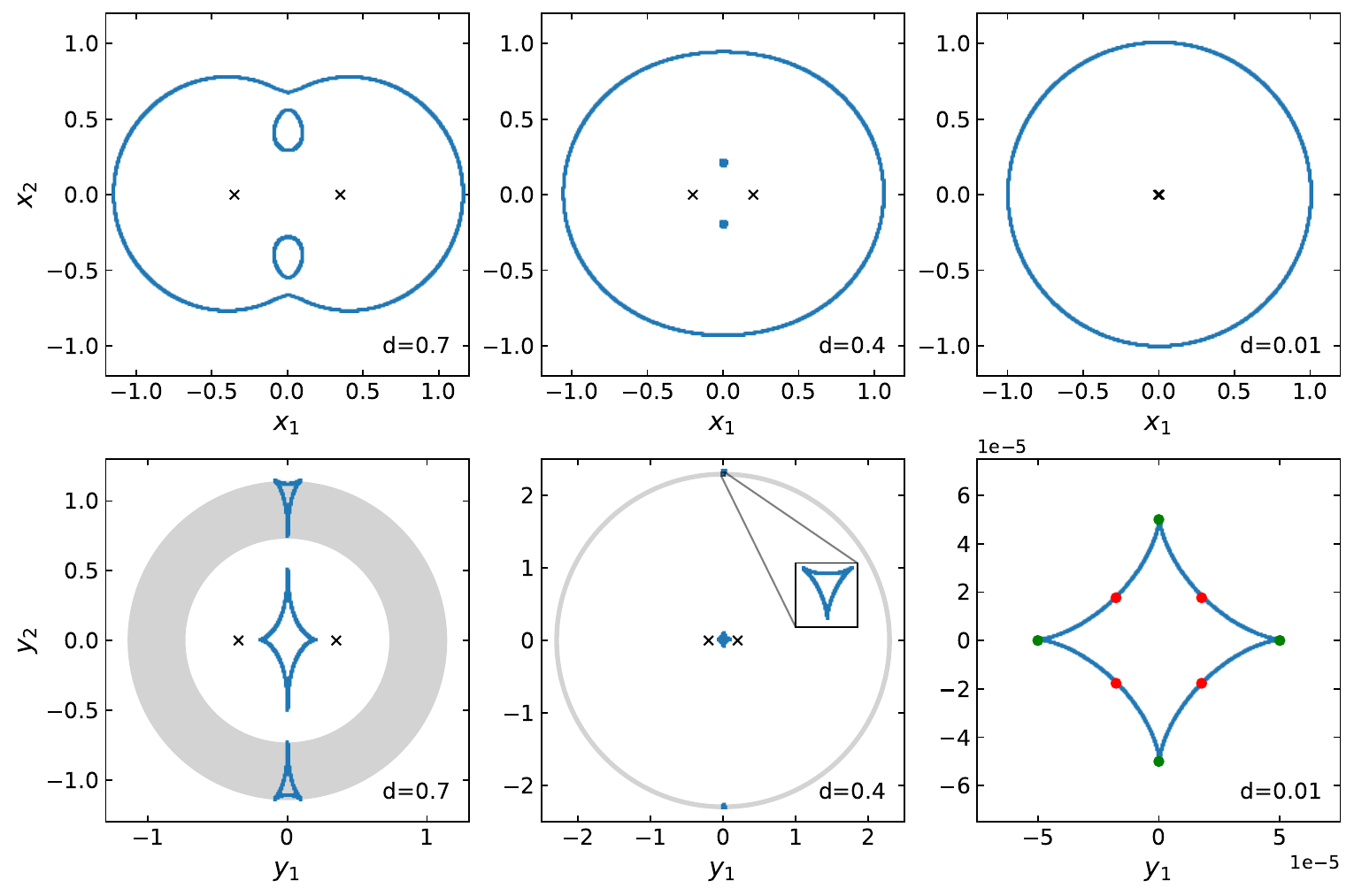}
    \caption{The critical curves and caustic curves for equal mass binary lenses and three values of $d<1$. $y_{1,2}$ and $x_{1,2}$ are the source and image position's coordinates in their respective planes, all in units of the Einstein radius (Eq.~\eqref{eq:ein_rad}). $q=1$ in this figure. The black crosses marked with $L_1$ and $L_2$ are the two SMBH lenses. In these cases, the critical and caustic curves have the same topologies independently of $d$: the critical curves form symmetrically two smaller closed curves enclosed by a larger curve, and the caustic curves are comprised of a central diamond-shaped curve and two outer triangular curves. The shaded region in the first column marks the region covered by the secondary (triangular) caustic during an orbital period (not shown in the right column). The inset in the caustic plot of the middle column shows a zoom-in of the upper secondary caustic curve. In the caustic plot of the right column, we mark the 4 cusp points with green dots and the 4 innermost fold points with red dots.\label{fig:caustic} }
\end{figure*}

\begin{figure*}
    \includegraphics[width=1\linewidth]{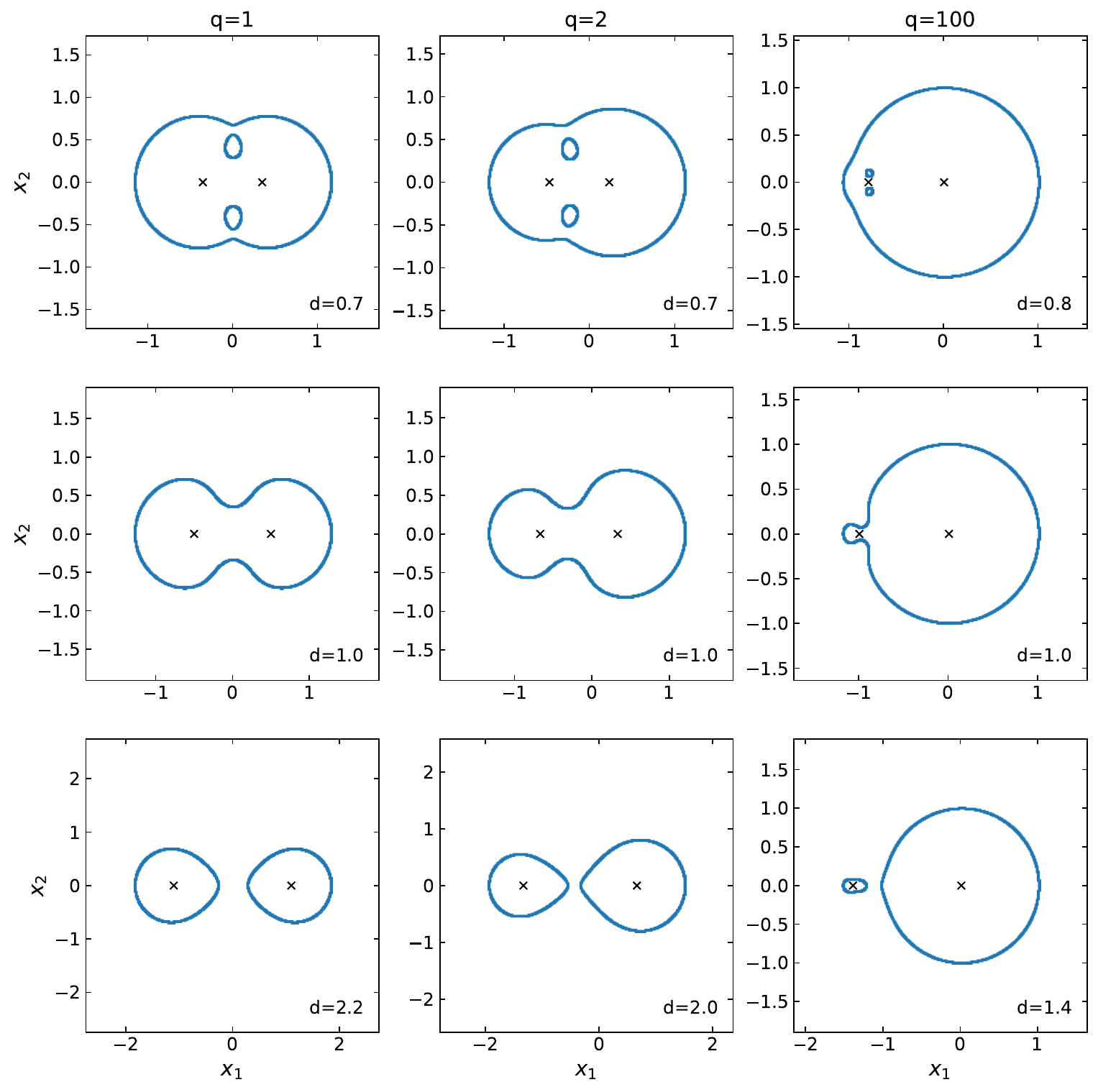}
    \caption{Example critical curves from all topologies for different mass ratios $q=m_1/m_2$. The black crosses mark the position of the two lenses. $x_{1,2}$ are the image positions in units of the Einstein radius. \label{fig:crit_all} } 
\end{figure*}

\begin{figure*}
    \includegraphics[width=1\linewidth]{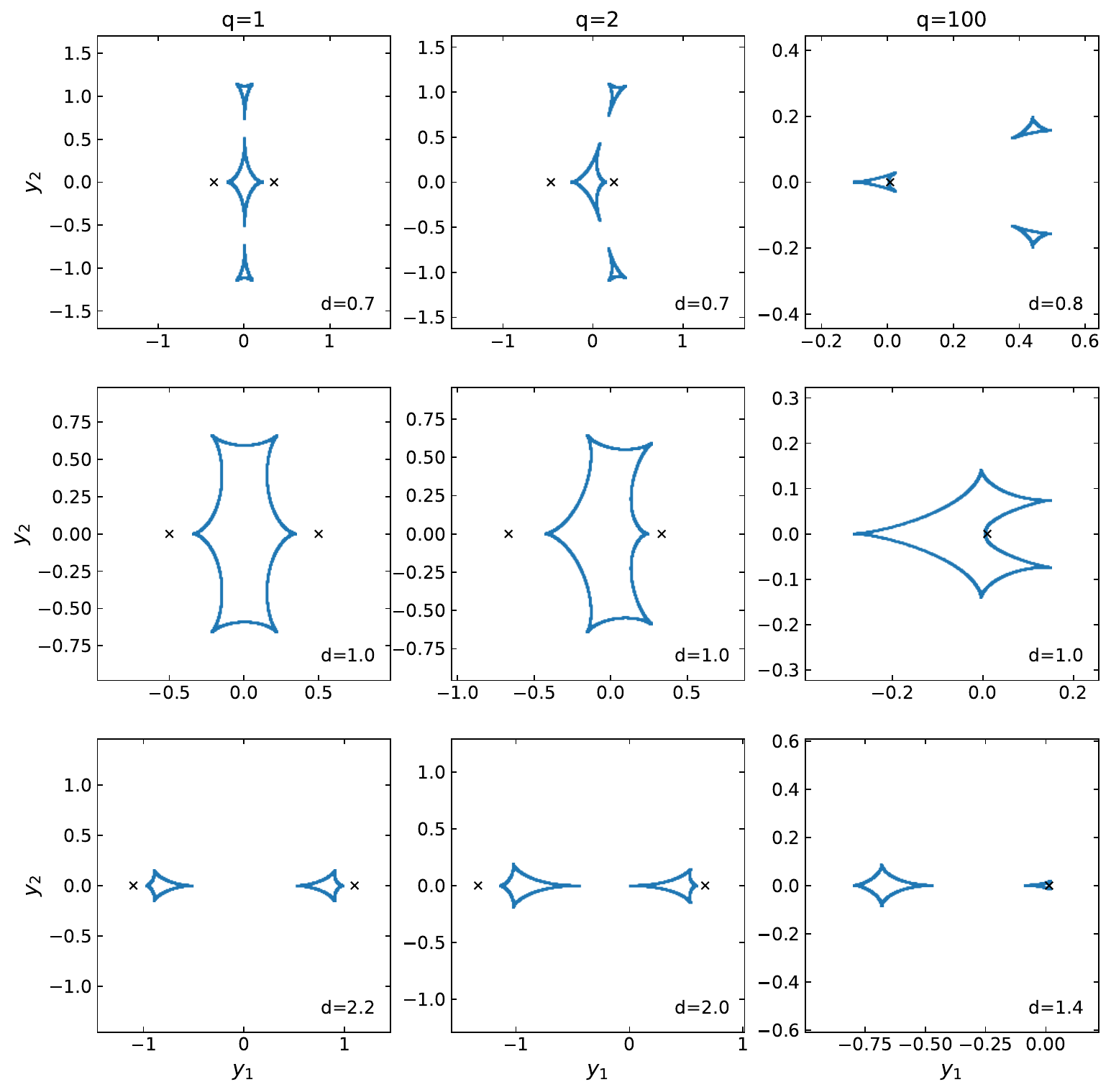}
    \caption{Example caustic curves from all topologies for different mass ratios $q=m_1/m_2$ corresponding to the critical curves in Fig.~\ref{fig:crit_all}. $y_{1,2}$ are the sources positions in units of the Einstein radius. The black crosses mark the position of the two lenses. Note that for the third column of $q=100$, we only mark the heavier lens position for better demonstration. Due to the large mass ratio, the caustic curve is much nearer the more massive lens.\label{fig:caustic_all}}
\end{figure*}

Fig.~\ref{fig:caustic} shows how the critical curves and caustic curves change as $d$ decreases toward small values in the close binary topology. As $d$ decreases across the columns, the outer primary critical curve moves closer to the Einstein ring, the inner secondary critical curves shrink in size and become closer to the center of mass. Both the central primary caustic curves and the secondary (triangular) caustic curves shrink, while the secondary caustic curves divert further from the center of mass. A source near the central primary caustic curve has highly magnified images near the corresponding primary 
critical curve, i.e. the Einstein ring. A source near the secondary caustic curve has magnified images near the corresponding secondary critical curves,near the center of mass. 
As the binary lens rotates, the secondary caustic curves span an annular area as seen in the shaded region of the first two columns of the caustic curve plots. The third column has $d\ll1$. In this case, we only plot the central primary 
caustic curve that has a diamond shape. The corresponding critical curve is similar to the Einstein ring. 
Figures~\ref{fig:crit_all} and~\ref{fig:caustic_all} show the effect of the mass ratio on the critical caustics and curves. 


\subsection{Tight Binary}

Eq.~\eqref{eq:crit_eqn} is a quadratic equation in terms of $z^2$. We study the solutions of $z$ with $|z|\approx1$, which form the critical curve that approximates the Einstein ring and the corresponding caustic curve that has a diamond shape as seen in Fig.~\ref{fig:caustic} for $d\ll1$. In this case $\pm \frac12 \theta$ approximately gives the azimuthal position on the Einstein ring. 

To quantify the caustic size, we focus on the innermost fold points (red dots in Fig.~\ref{fig:caustic}) and the cusp points (black dots in Fig.~\ref{fig:caustic}). They correspond to $\theta =(\frac12 \pi,\frac32 \pi)$ and $(0,\pi)$ respectively.

Expanding the solution of Eq.~\eqref{eq:crit_eqn} to $O(d^2)$ gives, the position of one innermost fold critical point with $\theta = \pi/2$ is at
\begin{equation}
    \label{eq:fold_pos_x}
    \bm{x}_{\rm fold} \approx \frac{1}{4\sqrt{2}}(4+3d^2,4-3d^2  ) + O(d^4).
\end{equation}
Mapping back to the source plane using the lens equation Eq.~\eqref{eq:leqn} gives the position of the fold caustic point marked by the bottom left red dot in Fig.~\ref{fig:caustic}
\begin{equation}
\label{eq:fold_pos_y}
    \bm{y}_{\rm fold}  \approx \frac{d^2}{4\sqrt{2}}(-1,-1)
\end{equation}
The position of the cusp critical point with $\theta = 0$ is at 
\begin{equation}
    \label{eq:cusp_pos_x}
    x_{\rm cusp} \approx 1+\frac{3d^2}{8}+O(d^4),
\end{equation}
along the $x_1$ axies.
The corresponding cusp caustic point marked by the right black dot is at 
\begin{equation}
    \label{eq:cusp_pos_y}
    y_{\rm cusp} \approx \frac{d^2}{2} +O(d^4),
\end{equation}
along the $y_1$ axis. Hence, the central caustic curve is between two circles centered on the center of mass of radius $d^2\xi_0/2$ and $d^2\xi_0/4$ respectively.

The secondary (triangular) caustic curves are at a distance of $\xi_0/d$ with an area of $\pi d^6 \xi_0^2/16$ \cite{Bozza:2000za}.

\subsection{Wide Binary}
For a wide binary with $d\gg1$,  the critical and caustic curves are approximately two Einstein rings and diamonds centered around the two lens. we solve Eq.~\eqref{eq:crit_eqn} for $A=\partial\bm{y}/\partial\bm{x}=0$ as well and expand around $1/d$, the caustic positions along the $y_1$ axis around $L_1$ are given by 
\begin{align}
    \label{eq:caus_wide_pos}
    y_{\rm cusp} \approx \frac{d}{2}-\frac{1}{2d}\pm\frac{1}{\sqrt{2}d^2} -\frac{3}{4d^3} +O(d^{-4}),
\end{align}
where the $\pm$ signs correspond to the two end points of the caustic curve along the $y_1$ axis. Taking their difference gives the diameter of the caustic curve to be approximately $\sqrt{2}/d^2$.

\section{Magnification of extended sources near the caustic curve} \label{app:mag}
In this section, we give estimate the
magnification of extended sources near caustic curves of different topologies.
It is convenient to work in a basis in which $\mathbf{A}$ (defined in Eq.~\ref{eq:A_def}) is diagonal which we shall denote with 
\begin{equation}
    \label{eq:phi_new}
    \tilde{\mathbf{A}} = \mathbf{O} \mathbf{A} \mathbf{O}^T.
\end{equation}
Here $\mathbf{O}$ is an orthogonal transformation that diagonalizes $\mathbf{A}$. Further, if $\mathbf{O}$ is applied to the source and image planes as $\tilde{\bm{y}} = \mathbf{O} \bm{y}$, $\tilde{\bm{x}} = \mathbf{O} \bm{x}$, and defining $\tilde{\psi}(\tilde{\bm{x}}) = \psi(\mathbf{O}^T\tilde{\bm{x}})$, it can be shown that the lens equation is preserved \cite{Petters:2001}:
\begin{equation}
    \label{eq:lens_trans}
    \tilde{\bm{y}} = \tilde{\bm{x}} - \tilde\nabla\tilde{\psi}(\tilde{\bm{x}}).
\end{equation}
Following Ref.~\cite{Petters:2001}, we expand the potential around a critical/caustic point which we choose to be the origin up to 3rd order:
\begin{equation}
    \label{eq:psi_exp}
    \psi(x_1,x_2) = E(x_1,x_2)+G(x_1,x_2)+H(x_1,x_2),
\end{equation}
where
\begin{align}
E(x_1,x_2) &=sx_1+tx_2, \nonumber \\
G(x_1,x_2) & = ax_1^2+bx_1x_2+cx_2^2, \nonumber \\
H(x_1,x_2) & = ex_1^3+fx_1^2x_2+gx_1x_2^2+hx_2^3,\label{eq:ABC_exp}
\end{align}
where $s,t,a,b,c,e,f,g,h$ are the Taylor expansion coefficients of the potential, e.g. $a=\psi_{11}/2$ and $b=\psi_{12}$. The subscripts of the potential denote derivative, e.g. $\psi_{12} = \partial_{x_2}\partial_{x_1}\psi(\bm{x})$, where $\bm{x}\equiv (x_1,x_2)$ are Cartesian coordinates.
The Jacobian matrix of the lens equation $\mathbf{A}=\partial\tilde{\bm{y}}/\partial\tilde{\bm{x}}$ is dominated by the 2$^{\rm nd}$ order terms:

\begin{equation}
    \label{eq:phi_mat} A=
\begin{pmatrix}
1-2a &-b \\
-b & 1-2c
\end{pmatrix}
-
\begin{pmatrix}
6e x_1 +2f x_2& 2fx_1 +2g x_2 \\
2fx_1 +2g x_2 & 6h x_2+2gx_1
\end{pmatrix}
.
\end{equation}
A critical point satisfies $\rm{det}(\mathbf{A})=0$. The transformation matrix at the critical point is given by
\begin{equation}
    \label{eq:O_mat}
    \mathbf{O} = \frac{1}{\sqrt{(1-2a)^2+b^2}}
    \begin{pmatrix}
        1-2a& -b \\
        b & 1-2a
    \end{pmatrix}
\end{equation}
and $\tilde{\mathbf{A}}(0,0) = \textrm{diag}(C,0)$, where
\begin{equation}
    \label{eq:K_def}
    C = 2-2(a+c).
\end{equation}
Given that $\psi$ (Eq.~\ref{eq:psi}) satisfies Poisson's equation in 2D, we have $\nabla^2 \psi(\bm{x})=2(a+c)=0$ for $\bm{x}_{m_1}\neq \bm{x}\neq\bm{x}_{m_2}$, hence $C=2$.

The expansion of the transformed lens equation \eqref{eq:lens_trans} up to 2nd order is
\begin{align}
    \tilde{y}_1(\tilde{x}_1,\tilde{x}_2) &=  C \tilde{x}_1 -2 \tilde{f} \tilde{x}_1 \tilde{x}_2 - \tilde{g}\tilde{x}_2^2, \nonumber \\
    \tilde{y}_2(\tilde{x}_1,\tilde{x}_2) &=-\tilde{f}\tilde{x}_1^2-2\tilde{g}\tilde{x}_1\tilde{x}_2-3\tilde{h}\tilde{x}_2^2. \label{eq:new_lens_eq}
\end{align}
The transformed expansion coefficients are given by, e.g. $\tilde{g}(\tilde{\bm{x}})=g(\mathbf{O}^T\tilde{\bm{x}})$. 
Assuming $1-2a\neq0$ without loss of generality, the fold points satisfy $\tilde{h} \neq0$ and the cusp points satisfy $\tilde{h}=0,\tilde{g}\neq0$ \cite{Schneider:1992}. Note that the $\tilde{x}_1^2$ term in $\tilde{y}_1$ is ignored because it is dominated by $\tilde{x}_1$. This allows us to approximate the caustic curve as a parabola by setting the Jacobian determinant of Eq.~\eqref{eq:new_lens_eq} to zero, $\mathrm{det} |\partial\tilde{\bm{y}}/\partial\tilde{\bm{x}}|=0$ and solve simultaneously with Eq.~\eqref{eq:new_lens_eq} \cite{Schneider:1992}. One side of the parabola has no images near the critical curve, the other side, i.e. the convex side, has two images. For a distance of $\Delta \tilde{y}_2$ from the $\tilde{y}_1$ axis, these two images both have to leading order an approximate magnification of 
\begin{equation}
    \label{eq:mag_approx_caustic}
    \mu = \frac{1}{2K\sqrt{|3\tilde{h}\Delta \tilde{y}_2|}} \equiv \frac{\mu_0}{\sqrt{\Delta \tilde{y}_2}},
\end{equation}
where $\mu_0$ is defined by this equation. For an extended source, the magnification in Eq~\eqref{eq:mag_approx_caustic} needs to be averaged over the source area. For uniform circular source with radius $R_{\rm src}= r\xi_0$ and distance $l \xi_0$ from the caustic, the magnification is \cite{Miralda-Escude1991}:
\begin{equation}
    \label{eq:mag_extend}
    \mu = \frac{\mu_0}{\sqrt{r}}F\left(\frac{l}{r}-1\right),
\end{equation}
where
\begin{align}
F(y_0) &= \frac{2}{\pi}\int_{-y_0}^{2} \mathrm{d}y \left[  \frac{y(2-y)}{y+y_0}\right]^{1/2} \qquad (y_0<0),  \nonumber \\ F(y_0) &= \frac{2}{\pi}\int_{0}^{2} \mathrm{d}y \left[  \frac{y(2-y)}{y+y_0}\right]^{1/2} \qquad (y_0>0). \label{eq:F_def_mag}
\end{align}
The maximum of $\mu$ in the case of extended sources occurs at $l = 0.65 r$, with \begin{equation}
    \label{eq:mu_max_extend}
    \mu_{\rm max} = 1.4\frac{\mu_0}{\sqrt{r}}
\end{equation}

Let us estimate the value of $\tilde{h}$ at the innermost fold point for $d\ll1$. Substituting Eq.~\eqref{eq:fold_pos_x} in the derivatives of the gravitational potential gives 
\begin{equation}
    \label{eq:hbar_approx_fold}
    |\bar{h}| \approx \frac{d^2}{8}+O(d^3).
\end{equation}
Counting both highly magnified images, we thus approximate the magnification near the fold as 
\begin{align}
    \label{eq:fold_mag_ext_theory}
    \mu_{\rm fold} &\approx 
    1.14 d^{-1}r^{-1/2} 
    =1.14 \frac{\xi_0}{a(1\pm e)} \sqrt{\frac{\xi_0}{R_{\rm src}}}
    \notag \\
    &=4\times10^{5}(1\pm e)^{-1}M_{10}^{5/12}
 T_{\rm yr}^{-2/3} 
 R_{10}^{-1/2}D_{\rm kpc}^{3/4}
\end{align}
where $a$ and $e$ are the semimajor axis and eccentricity of the SMBHB. The magnification of an extended source near the cusp is found empirically to follow
\begin{align}
    \label{eq:cusp_mag_ext_emp}
    \mu_{\rm cusp} &\approx 1.56 d^{-0.7} r^{-0.64}\notag \\ &=1.56 \left(\frac{a(1\pm e)}{\xi_0}\right)^{-0.7} \left(\frac{R_{\rm src}}{\xi_0} \right)^{-0.64}\notag \\ 
    &=1\times10^{6}(1\pm e)^{-0.7}M_{10}^{0.44}
 T_{\rm yr}^{-0.47} 
  R_{10}^{-0.64}D_{\rm kpc}^{0.67}.
\end{align}
We have $\mu_{\rm fold}/\mu_{\rm cusp} \approx (r/d^2)^{0.15}$. These results are valid only for sources whose size is smaller than the size of the caustic curve ($r<\frac12 d^2$ as seen in App.~\ref{app:caus_size}). In this regime, the cusp magnification is always higher than that of the fold but still of the same order.  

The magnification of the source at the secondary caustic curves are found numerically to be 
\begin{align}
    \label{eq:mag_sec}
    \mu_{\rm sec} &\approx 0.64 d^{1.63}r^{-0.6}\notag \\
     &=100 M_{10}^{0.03}T_{\rm \rm yr}^{1.1} 
 R_{10}^{-0.6}D_{\rm pc}^{-0.5} 
\end{align}

The maximum magnification of a source with radius $r$ in units of Einstein radii at the cusp of the caustic in the $d\gg1$ regime is found empirically to be \begin{align}
    \mu_{\rm wide}&\approx 1.4\,d^{0.54}r^{-0.66} \notag \\
    &=10^4M_{10}^{0.24}
 T_{\rm 100}^{0.36} 
 R_{10}^{-0.66}D_{\rm pc}^{0.06}
\end{align}
where $T_{100} = T/100$ yr and $D_{\rm pc} = D_{\rm LS}/$ pc. 

\section{Light curve minimum magnification estimate} \label{app:minimum_mag}
The trough of the magnification for circular-orbit cusp crossing occurs approximately at $(\pm \frac12 d^2,\pm \frac12 d^2)$, at $45\degree$, between two adjacent cusps. Because the source will be far away from the caustic curve, the magnification of an extend source is approximately the same as that of a point source \cite{Pejcha:2009}. 

There are three highly magnified images near a cusp inside the caustic curve with the magnification of \cite{Pejcha:2009}:
\begin{align}
    \mu^{(j)}_{\rm in} = \frac{1}{B_2} \sqrt{\frac{y_1}{y_1^3 -K y_2^2}} 
    \cos \left[ \frac{1}{3} \arcsin{\sqrt{\frac{Ky_2^2}{y_1^3}}+\frac{2}{3}\pi (j-1)}\right] 
   \label{eq:cusp_mag_inside}
\end{align}
where $j\in\{1,2,3\}$ and
\begin{equation}
    \label{eq:K_def_cusp}
    K = -\frac{27 B_3^2(2B_1B_3 +B_2^2)}{8B_2^3}.
\end{equation}
and
\begin{equation}
    \label{eq:cusp_coeff_def}
    B_1 = \frac{1}{6} \partial_{x_2}^4\psi, \,\,\, B_2= \partial_{x_2}^2\partial_{x_1}\psi,\,\,\, B_3= A_{11},
\end{equation}
where $A_{ij}=\delta_{ij}-\partial_j\partial_i\psi$ are the elements of the Jacobian matrix in Eq.~\eqref{eq:phi_mat}. Note that we have dropped the tildes on the expansion coefficients and $y_1,y_2$ for simplicity. All the expressions apply after transforming to the basis where the Jacobian matrix is orthogonal using Eq.~\eqref{eq:phi_new} and Eq.~\eqref{eq:O_mat}.

There is one highly magnified image outside the cusp with magnification of \cite{Pejcha:2009} 
\begin{align}
    &\mu_{\rm out} = -\frac{1}{2B_2K^{1/3}} 
    \frac{\left(u+y_2\right)^{1/3}+\left(u-y_2\right)^{1/3}}{u}.
    \label{eq:cusp_mag_outside}
\end{align}
where $u=(y_2^2- K^{-1}y_1^3)^{1/2}$. The
$(y_1,y_2)$ source position components
in Eqs.~\eqref{eq:cusp_mag_inside} and \eqref{eq:cusp_mag_outside} 
are defined in a coordinate system whose origin is at the nearby cusp point and the $y_1$ axis is aligned with the cusp axis of symmetry.

When the source is at $45\degree$ between two cusp points, there are two highly magnified images induced from both nearby cusp points. We thus need to include an additional factor of $2$ to Eq.~\eqref{eq:cusp_mag_outside}. At small $d$, we have $|K|\approx 27d^2/32$ and $|B_2|\approx2$. The minimum magnification between two nearby cusp crossings is estimated to be 
\begin{align}
    \label{eq:minmum_mag}
    \mu_{{\rm min},y=\frac12 d^2} &\approx 1.08 d^{-2} 
    \approx 1.9\cdot10^4 M_{10}^{1/3} T_{\rm yr}^{-4/3}D_{\rm kpc},
\end{align}
assuming that the source position is near the cusp radius, $y \approx \frac12 d^2$.

The magnification of a point-mass lens for a point source at impact parameter $y$ is given by 
\begin{equation}
    \label{eq:mag_pointmass}
    \mu_{\pm} = \frac{1}{2} \pm \frac{2+y^2}{2y\sqrt{4+y^2}},
\end{equation}
where $\pm$ stands for the two images with different parities. The observed magnitude is the sum of the absolute value of the two images: $\mu=|\mu_+|+|\mu_-|$. For $y\ll1$, the total magnification is approximately 
\begin{align}
    \label{eq:mag_point_approx}
    \mu_{\rm merger} \approx 
    \begin{cases}
        y^{-1} = 6\cdot10^4 M_{10}^{1/2} Y_{5}^{-1}D_{\rm kpc}^{1/2},  & \text{for } y > r\,,\\[2ex]
        2\,r^{-1} = 1.2\cdot 10^7\, M_{10}^{1/2} R_{10}^{-1} D_{\rm kpc}^{1/2} & \text{for } y < r\,.
    \end{cases}
\end{align}
where $y\xi_0$ is the impact parameter of the source and $Y_5=y\xi_0/5$ AU. This provides an estimate for the magnification near merger during \qpls{}.

\section{Lensing probability comparison of binary lenses and point-mass lenses}\label{app:Point_vs_binary}
In this section, we compare the strong lensing probability of a binary lens and a point-mass lens. We compute the area in the source plane in which the lensing magnification reach some high value $\mu_{\rm ref}$. 

For a binary lens with $d\ll1$, the extreme magnification lensing area is approximately the annulus bounded by the innermost fold points with radius $d^2/4$ and the cusp points with radius $d^2/2$. The peak magnification in this region is between $\mu_{\rm fold}$ (Eq.~\ref{eq:fold_mag_ext_theory}) and $\mu_{\rm cusp}$ (Eq.~\ref{eq:cusp_mag_ext_emp}). For the binary-lensing magnification to be larger than $\mu_{\rm ref}$, $d$ must be less than
\begin{equation}
    \label{eq:d_mu_ref}
    d_{\rm BL}=1.14 \, r^{-1/2}\mu_{\rm ref}^{-1},
\end{equation}
and the corresponding area in the source plane (in units of $\xi_0^2$) is
\begin{equation}
    \label{eq:lensing_area_BL} 
    A_{\rm BL} =  \frac{3\pi d_{\rm BL}^4}{16} = \frac1{r^{2}\mu_{\rm ref}^{4}}.
\end{equation}

For the point-mass lens, we approximate the magnification close to the lens using Eq.~\eqref{eq:mag_point_approx}. Within the circle with radius $y=\mu_{\rm ref}^{-1}$, the magnification is larger than $\mu_{\rm ref}$. Hence the lensing area of the point-mass lens (in the same units of $\xi_0$ as for the binary lens with the same total mass)  is 
\begin{equation}
    \label{eq:lensing_area_PL}
    A_{\rm PL} = \frac{\pi}{\mu_{\rm ref}^2}.
\end{equation}

The ratio of the lensing areas gives the relative probability of binary lensing to point-mass lensing as
\begin{align}
    \label{eq:lensing_ratio}
   \frac{A_{\rm BL}}{A_{\rm PL}} &= \frac{1}{\pi r^{2} \mu_{\rm ref}^{2}} 
   = \frac{0.12\, a^2(1+ e)^2}{R_{\rm src}\sqrt{R_{\rm g} D_{\rm LS}}}
   \notag \\&
   \approx 86 (1+e)^2 M_{10}^{1/6} T_{\rm yr}^{4/3} R_{10}^{-1} D_{\rm kpc}^{-1/2}
\end{align}
where $R_g=GM/c^2$ is the gravitational radius. The probability of high magnification by binaries is much higher than singles, especially for wide binaries and spatially compact sources.

\section{Caustic crossing time}\label{app:crossing_time}
If the caustic revolves at the instantaneous SMBHB angular frequency $\omega$ during caustic crossing, the caustic crossing time for a caustic of size 
\begin{equation}\label{eq:Rc}
R_c=\frac12 \xi_0 d^2 = \frac12 \frac{(1\pm e)^2a^2}{\xi_0} = \frac14 \frac{(1\pm e)^2a^2}{\sqrt{R_{\rm g}D_{\rm LS}}}
\end{equation}
by an object of diameter $2R_{\rm src}$ at the periapsis and apoapsis is 
\begin{align}\label{eq:caus_cross_time_detailed}
t_{\rm mag} &= \frac{2 R_{\rm src}}{\omega R_c} =
\frac{4R_{\rm src} \xi_0}{\omega R_{\rm sep}^2} = \frac{4R_{\rm src} \xi_0}{\sqrt{GMa(1-e^2)}}
\notag\\
&= 2T_{\rm orb}\frac{R_{\rm src} \xi_0}{\pi a^2 \sqrt{1-e^2}}
\notag\\
&=16\,{\rm hr}\,\frac{R_{10} {M_{10}}^{-1/6} T_{\rm yr}^{-1/3} D_{\rm kpc}^{1/2}}{\sqrt{1-e^2}}
\end{align}
In the second equality we used $d=R_{\rm sep}/\xi_0$, and in the third we identified the specific angular momentum with $\omega R_{\rm sep}^2 = \sqrt{GMa(1-e^2)}$ and noted that it is a constant of motion. In this calculation we neglect the change of the SMBHB separation, valid at apoapsis and periapsis. 
The result is equal to twice the orbital time times a geometrical factor: the ratio of the source radius times Einstein radius over the area of the SMBH orbit. 

The crossing time of the secondary caustic (located at radius $\xi_0/d$ from the center of mass) can be obtained similarly as Eq.~\eqref{eq:caus_cross_time_detailed}
\begin{align}
    \label{eq:caus_cross_sec}
    t_{\rm mag,sec} &= \frac{2R_{\rm src}}{\omega \xi_0/d} \notag \\
    &= 12  {\rm s} \, M_{10}^{-2/3}T_{\rm yr}^{5/3} 
 R_{10}D_{\rm pc}^{-1} \frac{(1\pm e)^3}{\sqrt{1-e^2}}
\end{align}

\section{Magnification outside of the caustic}
\label{sec:outside_caustic}
If the source is placed further from the caustic curve, we can still get a light curve, at a smaller magnification. For example, we place the source at $y=d^2$, hence at double the caustic radius from the center of mass.

The maximum magnification is given when the source is aligned with one of the cusp axis, we add the magnification from Eq.~\eqref{eq:cusp_mag_inside} of the three nearby cusp points to estimate the magnification to be ($y=d^2$)
\begin{align}
    \label{eq:max_mag_far}
        \mu_{{\rm far,max}, y=d^2} &\approx 1.6 d^{-2} 
        \approx 2.8\cdot10^4 M_{10}^{1/3} T_{\rm yr}^{-4/3}D_{\rm kpc},
\end{align}

The minimum magnification is again given when the source is at $45\degree$, between two cusp points. We add up the contribution from the nearest two cusp points and estimate the minimum magnification to be 
\begin{align}
    \label{eq:min_mag_far}
        \mu_{\rm far,min} &\approx 0.98d^{-2} \notag \\
    & \approx 1.7\cdot10^4 M_{10}^{1/3} T_{\rm yr}^{-4/3}D_{\rm kpc},
\end{align}
Eqs.~\eqref{eq:max_mag_far} and \eqref{eq:min_mag_far} are all verified numerically.

\begin{figure*}
    \includegraphics[width=1\linewidth]{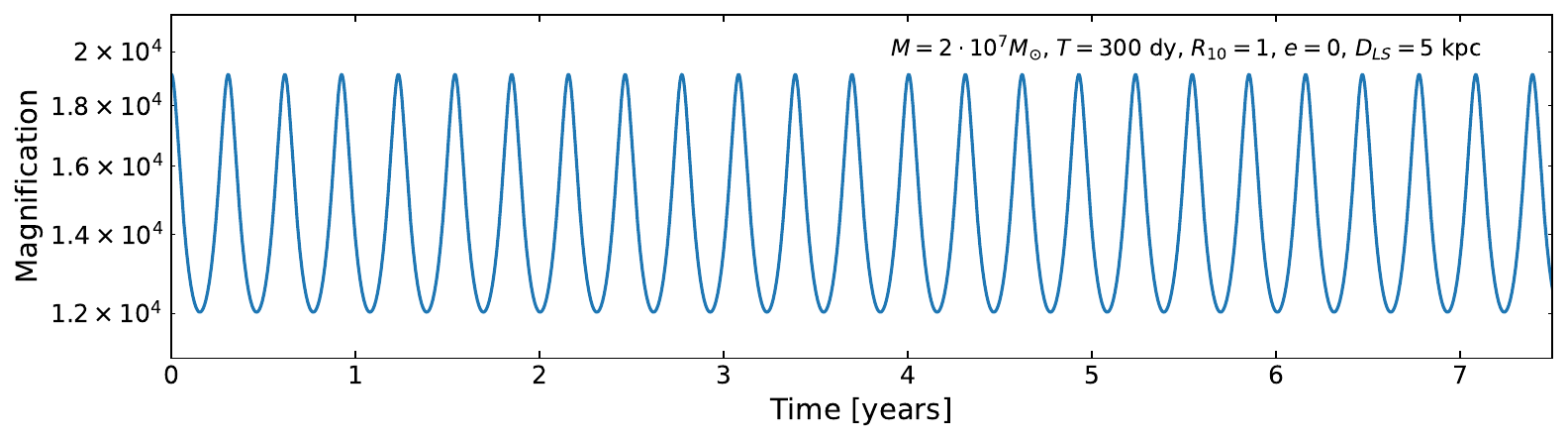}
    \caption{Same parameter as the first panel of Fig.~\ref{fig:lisa_light_curve}, but the source is now placed at double the caustic size distance away from the center of mass. The host galaxy is assumed to be at redshift $z=0.5$.}\label{fig:far_source_light_curve}
\end{figure*}

\section{Detection thresholds}\label{sec:detection_thresholds}

The high magnification achieved by SMBHBs allows individual stars to produce a potentially observable imprint. 
One can write the apparent magnitude of the lensed star as
\begin{align}
    m =& 44.83 -2.5\log_{10}\left(\frac{\mu L_*}{L_\odot}\right) + 5 \log_{10}\left(\frac{D_L}{\rm Gpc}\right)  \\
    = & 30.82 -2.5\log_{10}\left(M_{10}^{5/12}T_{\rm yr}^{-2/3}R_{10}^{-1/2}D_{\rm kpc}^{3/4}\right) \nonumber \\
    & -2.5\log_{10}\left(\frac{L_*}{L_\odot}\right) + 5 \log_{10}\left(\frac{D_L(1+z)^2}{\rm Gpc}\right)\,,
\end{align}
where the absolute solar magnitude of 4.83 was used and the last line replaces $\mu$ with Eq.~(\ref{eq:mag_ext_fold_withunits}) with $e=0$.

Solving for $D_{\rm L}$ gives the maximum distance of detection at a given $m$ apparent magnitude of
\begin{align}
    \label{eq:Dl_horizon}
    D_L(1+z)^2=& 3.4\, {\rm Gpc} \cdot 10^{(m-26)/5} M_{10}^{5/24}T_{\rm yr}^{-1/3} \notag \\
    &\cdot R_{10}^{-1/4} {L_{1000}^{1/2}}D_{\rm kpc}^{3/8}
\end{align}
where the source absolute luminosity is $L_{1000}=L_*/(10^3L_{\odot})$.

\begin{figure}
    \centering
    \includegraphics[width=\linewidth]{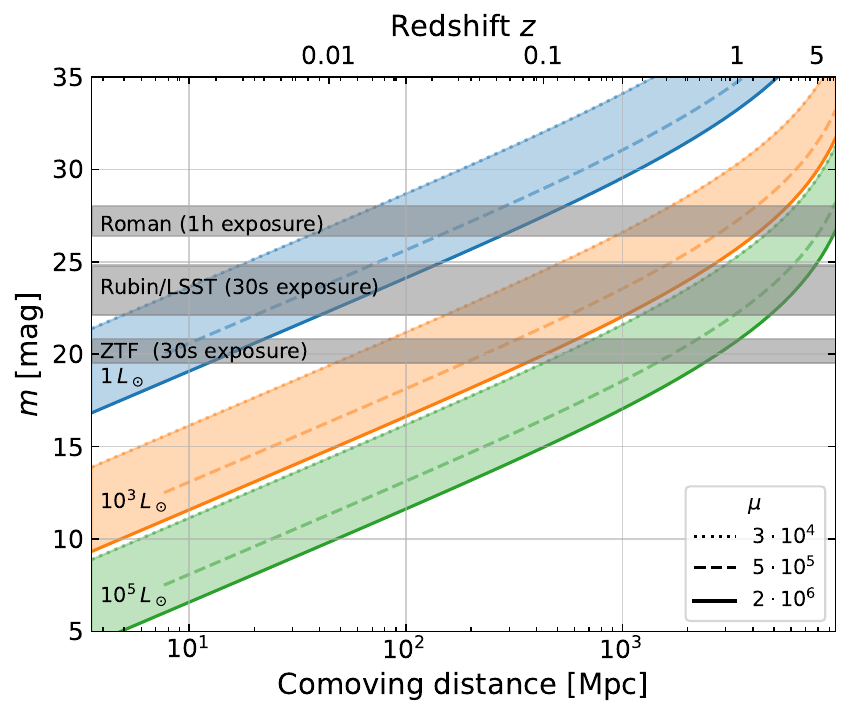}
    \caption{Detectability of stars lensed by SMBHBs. Bands show the lensed magnitudes as a function of distance for stars of 1, $10^3$ and $10^5$ solar luminosities and different level of magnifications representative of Fig.~\ref{fig:massive_light_curve}. Horizontal bars show the range of single-exposure sensitivities for the filters in ZTF, Rubin-LSST and Roman.}
    \label{fig:detetctability_thresholds}
\end{figure}

Fig.~\ref{fig:detetctability_thresholds} shows the magnitude of lensed stars of different intrinsic luminosities as a function of the source's distance. The intrinsic luminosities represent different stellar types and the range of $\mu$ is based on the lighcurve in  Fig.~\ref{fig:massive_light_curve}, with lines corresponding to the minimum (dotted), minimum between folds (dashed) and cusp crossing (solid).

Horizontal bars show the single-exposure and single-filter detection thresholds for ZTF~\cite{Bellm_ZTF}, Rubin/LSST~\cite{2022_LSST_Cadence_Bianco} and Roman~\cite{roman_wfi_technical} (5-$\sigma$ detection for point sources), indicative of the distance at which a given magnified source can be identified. These triggers are derived for point sources: a more detailed estimate needs to address the variability in the crowded field of a galactic center, as well as the gain in sensitivity from following the lightcurve over many exposures and several filters.

The effect of the source size on the peak magnification can partially compensate for the lower brightness of smaller stars. Assuming the setup of Fig.~\ref{fig:massive_light_curve}, a Sun-like star with $L_* = L_\odot, R_{*}\sim R_\odot$ can be detected by Rubin-LSST (Roman) at a distance $\lesssim 80 \;(400)\,\text{Mpc}$
based on single exposure sensitivity of 24 (27.63) mag at 750nm (see Eq.~\eqref{eq:Dl_horizon}). 
A young white dwarf with $L_*= 0.01 L_\odot$, $R_{*}\sim 0.01 R_{\odot}$ can be detected at $\lesssim 24 \;(130)\,\text{Mpc}$
under similar detection thresholds, but has a shorter duration of the magnification peak $\sim 50\, {\rm s} \,M_{10}^{-1/6}  T_{\rm  yr}^{-1/3} (R_{10}/10^{-3})  D_{\rm kpc}^{1/2}$ at each caustic crossing, cf.~Eq.~(6).

\section{GW-driven binary evolution}\label{app:time_to_merge}

The evolution of SMBHBs is dominated by GW emission at sufficiently small separations. The energy loss due to radiated orbital energy causes the binary's frequency to increase. The time spent as a function of the frequency is given by 
\begin{equation}\label{eq:dt_df}
    \frac{dt}{df} = \frac{5c^5}{96 \pi^{8/3}}  \left(G\mathcal{M}\right)^{-5/3} f^{-11/3}\,, 
\end{equation}
where $t,f$ are source-frame quantities, $\mathcal{M} = M \left( \frac{q}{(1+q)^2} \right)^{3/5}$ is the source-frame chirp mass and a quasi-circular binary has been assumed. Converting to detector-frame quantities introduces a factor $(1+z)^{-2}$ to account for time dilation. 
This relation also relates the total number of sources in a frequency range to the merger rate~\cite{Sesana:2008}
\begin{equation}\label{eq:dN_df}
    \frac{dN}{dz df} = \frac{dn}{dz dt}\frac{dt}{df}\,,
\end{equation}
where $\frac{dn}{dz}$ is the merger rate per unit redshift. Eq.~(\ref{eq:dt_df}) ceases to be valid at larger separations, when other effects (hardening due to gas or stellar scattering, etc.) become dominant. Therefore, estimates for the total number of sources need to replace $dt/df$ with the timescale associated to the dominant astrophysical process or for a very conservative estimate, limit the integration range in Eq.~(\ref{eq:dN_df}) to the lowest frequency where GW-driven evolution dominates the dynamics.

Eq.~(\ref{eq:dt_df}) can be used to compute the time to merger. Including orbital eccentricity, it is given by \cite{GW_review}
\begin{equation}
    \label{eq:merger} 
    t_{\rm merge} = \frac{5c^5}{256 G^3} g(e) (1-e^2)^{7/2} \frac{a^4 (1+q)^2}{M^3q},
\end{equation}
where $g(e)$ is a weak function of the eccentricity $e$. $g(e)$ can be analytically fitted as  \cite{Mandel:2021} 
\begin{equation}
    \label{eq:g_e_approx}
    g(e) \approx \left(1+0.27 e^{10} + 0.33 e^{20} + 0.2 e^{1000}\right) 
\end{equation}
In this work we will consider the leading order effect of eccentricity on the evolution of the SMBHB orbital frequency. We will ignore other effects, including higher GW emission and periastron precession.

For a highly eccentric SMBHB, we can observe quasi-periodic lensing signatures while the GW frequency of the SMBHB is already in the LISA band. The GW emission is dominated at the pericenter passage with a burst frequency $f_{\rm burst}$ much larger than the orbital frequency $f_{\rm orb}$ \cite{Xuan2024}: 
\begin{align}
\label{eq:fburst}
    f_{\rm burst} &\sim 2f_{\rm orb }(1-e)^{-3/2} \notag \\ &= 10^{-4}{\mathrm{\,Hz}} \left(\frac{T}{80 \,{\rm day}}\right)^{-1}\left(\frac{1-e}{0.02}\right)^{-3/2}
\end{align}
where $T=1/f_{\rm orb}$ is the orbital period.
The SNR of a single burst is estimated to be \cite{Xuan2024} 
\begin{align}
    \label{eq:SNR_burst}
{\rm SNR} &\sim \frac{h_{\rm burst}}{\sqrt{f_{\rm burst,z}S_n(f_{\rm burst,z}) }}\notag \\
&\approx 10  \left(\frac{M_z}{8\cdot10^6 M_{\odot}}\right)^{5/3} \left(\frac{T_z}{\rm 80 \, day}\right)^{-1/6} \notag\\
&\cdot\left(\frac{1-e}{0.02}\right)^{-1/4} \left(\frac{D_L(1+z)^2}{\rm3\, Gpc}\right)^{-1}\left(\frac{S_n(f_{{\rm burst},z})}{S_n(10^{-4}{\rm \, Hz})}\right),
\end{align}
where $M_z=M(1+z)$, $T_z=T(1+z)$, $f_{z}=f/(1+z)$ and $S_n(f_z)$ is the spectral noise density of LISA \cite{Babak2021}. 

The SNR of the entire duration of the GW emitted by the initially eccentric SMBHB before merger is estimated using the python package LEGWORK \cite{LEGWORK_apjs} \cite{LEGWORK_joss}. The package considers the evolution of the GW source in the inspiral phase and sums up the contributions from different harmonics for eccentric sources. For the example of highly eccentric ($e=0.98$) SMBHB presented in the main text (bottom panel of Fig.~\ref{fig:lisa_light_curve}), we computed the SNR to be 10 at $z\approx4.9$.

\section{\qpls{} and caustic crossing rates}\label{sec:probability}

We will estimate detection probabilities in two limits: moving stars crossing a static caustic and the volume covered by the caustic network of a rotating binary. The results depend strongly on the number density of bright stars. We will quote our results for a fiducial stellar density of $n_\star = 1 {\rm pc}^{-3}$: the resulting rates can be rescaled for different populations. We will comment discuss reasonable values of $n_\star$ below.

We will focus on equal mass ($q=1$), quasi-circular ($e=0$) and close-by $a<10^3 R_{\rm g}$ binaries, where the dimensionless separation is small
$d\ll 1$ (see Eq.~\ref{eq:d_dimensionless_scale} and Fig.~\ref{fig:caustic_size}).
In this limit there is a central caustic with radius $R_c\approx \frac12 \xi_0 d^2$, plus two secondary (triangular) caustics located at distance $R_2\approx \xi_0/d$ from the center, each with size $\delta R_2\approx \xi_0 d^3$~\cite[Eq.~45]{Bozza:2000za} (see Fig.~\ref{fig:caustic}).Separate caustic regions are connected by a region where the magnification is high, but finite~\cite[Fig.~1]{Penny:2011}. 

We will first compute the probability of \qpls{} and caustic-crossing rates for head-on SMBHBs. We will then consider the projection effects, before estimating the total rates from synthetic SMBHB catalogs.

\subsection{\qpls{} probability}\label{app:qpls_prob}

The expected number of stars producing \qpls{} events for a single SMBHB, or roughly the \qpls{} probability, is determined by the volume of the primary caustic and the velocity distribution of sources. We will estimate it as the 
\begin{equation}\label{eq:qpls_prob_base}
\mathcal{N}_{\rm QPLS} = \int_{D_{\rm tight}}^\infty \mathrm{d}D_{\rm LS}\, n_\star \pi R_c^2 P(v<v_0)\,,
\end{equation}
where \change{$R_c$ is given by Eq.~\eqref{eq:Rc} and} $D_{\rm tight} = a^2/(4R_g) = 0.06 \, {\rm pc} \, T_{\rm yr}^{4/3} M_{10}^{-1/3}$ enforces the tight-binary limit $d\leq 1$. The integral is weighted by the probability of the velocity being below
\begin{equation}\label{eq:v0}
v_0 = \frac{2R_c}{N_{\rm orb} T} \approx
\frac{a^2}{N_{\rm orb} \xi_0 T}
\approx 77\,\frac{\rm km}{\rm s}\, \frac{T_{\rm yr}^{1/3} M_{10}^{1/6}}{D_{\rm kpc}^{1/2} N_{\rm orb}}\,,
\end{equation}
such that the source takes at least $N_{\rm orb}$ binary periods to cross the caustic diameter \change{before drifting out of the caustic region}. We will assume the velocity distribution to be given by a Gaussian with zero mean and one-dimensional dispersion $\sigma_{\rm tot}$, so
\begin{align}
    P(v < v_0) &= \frac{1}{2\pi \sigma_{\rm tot}^2}\int_{0}^{v_0} \mathrm{d}v \, 2\pi v\, e^{-v^2/(2\sigma_{\rm tot}^2)}\notag\\
    &=\left(1-e^{-v_0^2/(2\sigma_{\rm tot}^2)}\right) \label{eq:probability_slow}
\end{align}
The total velocity dispersion, $\sigma_{\rm tot}=\sqrt{\sigma_v^2 + \sigma_{v,M}^2}$, where the contribution from the galactic potential \change{$\sigma_v$} is given by  (Eq.~\ref{eq:M_sigma}) and the contribution from the SMBHB follows from the virial theorem $\sigma_{v,M} = ({GM}/{r})^{1/2} = 207 \,{\rm km}\,{\rm s}^{-1}M_{10}^{1/2} D_{\rm kpc}^{-1/2}$. The distance at which $\sigma_v=\sigma_{v,M}$ defines the sphere of influence of the SMBHB $D_{\rm infl} = 0.23 \, {\rm kpc} \, M_{10}^{0.54}$. \change{The total velocity dispersion is dominated by $\sigma_v$ outside the sphere of influence and dominated by $\sigma_{v,M}$ inside the sphere of influence.}

\change{Well within the sphere of influence, the probability of sufficiently slow sources to allow caustic crossings for at least $N_{\rm orb}$ number of SMBHB periods, $P(v<v_0)\approx 1-\exp(-v_0^2/(2\sigma_{v,M}^2))$ in Eq.~\eqref{app:qpls_prob}, is independent of $D_{\rm LS}$,  given that\footnote{\change{here we assume circular SMBHBs and neglect a factor of $(1\pm e)^2$}}
 \begin{align}\label{eq:slow_cond_sphere_infl}
 \frac{v_{0}}{\sigma_{v,M}} &= \frac{2 R_c}{N_{\rm orb} T}\sqrt{\frac{D_{\rm LS}}{c^2 R_g}} =  
  \frac{a^2}{2c N_{\rm orb} R_{\rm g}T} \nonumber\\
 &=
 0.37\, T_{\rm yr}^{1/3}M_{10}^{-1/3} N_{\rm orb}^{-1}\,.
 \end{align}
Well outside the sphere of influence, $P(v<v_0)$ decreases as $D_{\rm LS}$ increases. We define a maximum $D_{\rm LS}$ distance 
\begin{equation}
D_{\rm cross} = \frac{a^4}{4 R_g \sigma_v^2 T^2 N_{\rm orb}^2} = 31 \, {\rm pc} \, T_{\rm yr}^{2/3} M_{10}^{-0.12} N_{\rm orb}^{-2}\,.    
\end{equation}
such that $v_0/\sigma_v\geq1$ for $D_{\rm infl}<D_{\rm LS}<D_{\rm cross}$, implying that $P(v<v_0)\approx1$. The integral in Eq.~\eqref{eq:qpls_prob_base} is dominated by $D_{\rm LS}<D_{\rm cross}$ so it may be truncated at the upper limit at $D_{\rm cross}$ and within this region we set $P(v<v_0)\approx1$.}

\change{Thus, given that $P(v<v_0)$ is approximately independent of $D_{\rm LS}$ both well inside and well outside the sphere of influence, it may be approximated by a constant 
\begin{equation}\label{eq:P0_values}
P_0 = 
\begin{cases}
     1, & D_{\rm infl}>D_{\rm cross},  \\
     1-\exp\left( -\frac{0.07  T_{\rm yr}^{2/3}}{M_{10}^{2/3} N_{\rm orb}^{2}}\right), &D_{\rm infl}\leq D_{\rm cross},
\end{cases}
\end{equation}
and taken outside of the integral in Eq.~\eqref{eq:qpls_prob_base}.
The expected number of \qpls{} sources which persist for at least $N_{\rm orb}$ binary periods without the source moving out of the caustic region, can be expressed approximately as}
  \begin{align}
  \label{eq:Nqpls_p0_approx}
      \mathcal{N}_{\rm QPLS} &\approx 
      \frac{\pi}{4} P_0 \int_{D_{\rm tight}}^{\tilde D}\mathrm{d}D_{\rm LS} n_\star  \xi_0^2 d^4 
      \nonumber
      \\ &= 4.73\cdot 10^{-6} \left(\frac{n_\star}{{\rm pc}^{-3}}\right) P_0  T_{\rm yr}^{8/3}M_{10}^{1/3}\ln\left(\frac{\tilde D}{D_{\rm tight}}\right)\,,
 \end{align}
\change{ where $\tilde D = \max(D_{\rm infl},D_{\rm cross})$.\footnote{\change{If $P_0(v<v_0)$ is not approximated as a constant in Eq.~\eqref{eq:qpls_prob_base}, the integral may still be approximately evaluated analytically in Eq.~\eqref{eq:Nqpls_p0_approx} by substituting the limiting cases for Eq.~\eqref{eq:probability_slow} from $D_{\rm tight}$ to $D_{\rm infl}$ and $D_{\rm infl}$ to $D_{\rm cross}$ if $D_{\rm infl}<D_{\rm cross} $ with different $P_0$ values and logarithmic factors. We use a single $P_0$ value and a single logarithmic factor in the final expression for simplicity, which is a good approximation, especially if $D_{\rm infl} \gg D_{\rm cross}$ (more massive and shorter period binaries) or $D_{\rm infl} \ll D_{\rm cross}$ (less massive and longer period binaries).}}} Substituting typical values in the logarithm yields
 \begin{equation}\label{eq:qpls_prob}
     \mathcal{N}_{\rm QPLS} = 4.0 \cdot 10^{-5} P_0 \left(\frac{n_\star}{{\rm pc}^{-3}}\right) T_{\rm yr}^{8/3} M_{10}^{1/3}F_Q\,.
 \end{equation}
 where the logarithmic correction is
$F_Q = 1+0.11\ln(T_{\rm yr}^{-4/3}M_{10}^{0.88}) $ for $(D_{\rm infl}>D_{\rm cross})$ and $0.75\big(1+0.16\ln(T_{\rm yr}^{-2/3}M_{10}^{0.23}N_{\rm orb}^{-2})\big)$ otherwise. Despite the crude assumptions, Eq.~\eqref{eq:qpls_prob} reproduces Eq.~\eqref{eq:qpls_prob_base} within 15\% accuracy for $M>10^5 M_\odot$ and $T<100$ yr. 

For a singular isothermal sphere (SIS) distribution for the stellar density follows $n_*(R) = N_{\rm c, pc}/(4\pi R^2 \mathrm{pc})$, where $N_{\rm c, pc}$ is the enclosed number of stars in the galactic center within 1 pc. The number of \qpls{} sources in the host galaxy of a SMBHB is 

\begin{align}
    \label{eq:N_qpls_SIS}
   \mathcal{N}_{\rm QPLS}^{\rm iso}=&\pi\int_{D_{\rm tight}}^{\tilde{D}} \mathrm{d}D_{\rm LS} \left(\frac12\xi_0 d^2\right)^2 n_*(D_{\rm LS}) P_0 \notag \\
   &=\frac{\ a^4 N_{\rm c, pc}}{32 R_g \, {\rm pc}}\left(D_{\rm tight}^{-2}-\tilde{D}^{-2}\right)P_0 \notag \\
   & \approx  \frac{ {N_{\rm c, pc}}R_g}{4\rm pc} P_0= 0.24 \left(\frac{N_{\rm c, pc}}{10^3}\right)\, M_{10} P_0,
\end{align}
where we have assumed that $\tilde{D} \gg D_{\min}$ in the last line. Note that this is an optimistic estimation because the stars in the galactic nucleus can be ejected due the presence of the SMBHB, resulting in a 'mass deficit' \cite{Merritt:2006}.

\subsection{Caustic-crossing rates per binary}\label{app:caustic_crossing_rate}

We will estimate the \qpls{} rates based on the caustic crossings in three different situations:  
\begin{itemize}
    \item[$\mathcal{R}_0$:] Stars with given velocity crossing the central caustic region, 
    \item[$\mathcal{R}_{c1}$:] Stars within the central caustic region of an orbiting binary,
    \item[$\mathcal{R}_{c2}$:] Rotation of the secondary caustic caustic region due to the binary's orbit. 
\end{itemize}

We will first compute the individual rates for a head-on binary. At the end of the section we consider the effect random orientations of the orbital plane. 
The rates-per-binary will be given in the source frame. Cosmological time dilation will be included in the next subsection, when we estimate total rates from simulated SMBHB catalogs. 

Let us first consider the case of a star with constant velocity entering the central caustic region, which we assume to be static. The rate is given by 
\begin{equation}
\mathcal{R}_0 \sim \langle \mathcal{S}_c  n_\star v_\star \rangle\,,    
\end{equation}
i.e.~the product of the caustic's area $(\mathcal{S}_c)$, stellar density $(n_\star)$ and velocity $(v_\star)$, averaged over all possible configurations. Note that this only describes the probability of stars entering the caustic region of the lens: this situation will be associated with several magnification peaks in general.

For $d\ll 1$, we approximate the caustic area by the cylinder of physical radius $R_c \approx \frac12\xi_0d^2$, so $\mathcal{S}_c = 2\pi \int_0^{\tilde D} \mathrm{d}D_{\rm LS}\,  R_c$, 
where $\tilde D$ is the 
maximum distance from the lens (e.g. the size of the host galaxy or where $d=1$). The area converges to 
\begin{equation}
    \mathcal{S}_c \approx 
    \pi a^2 \sqrt{\frac{\tilde{D}}{R_{\rm g}}} 
    =  0.5 \, {\rm pc}^2\, T_{\rm yr}^{4/3}(M_{10})^{1/6} \tilde D_{\rm kpc}^{1/2}\,,
\end{equation}
where $\tilde D_{\rm kpc}=\tilde D/1{\rm kpc}$ and $d\ll 1$ corresponds to $\tilde D \gg 0.06 \; {\text pc} \; M_{10}^{-1/3}T_{\rm yr}^{4/3}$ (see Eq.~\ref{eq:d_dimensionless_scale}).
Here for simplicity, we will assume a homogeneous density of bright stars $n_\star$ and mean velocity given by Eq.~(\ref{eq:M_sigma}): $\langle \mathcal{S}_c n_\star v_\star \rangle \approx \mathcal{S}_c \sigma_v n_\star$. 
The rate of caustic crossings by moving stars in the source frame is then
\begin{equation}\label{eq:rate_static}
 \mathcal{R}_0  \approx 2.2
 \cdot 10^{-4}\text{yr}^{-1} \left(\frac{n_\star}{{\rm pc}^{-3}}\right)T_{\rm yr}^{4/3}M_{10}^{0.39} \tilde D_{\rm kpc}^{1/2} \,.
\end{equation}
Note that we have only considered the velocity dispersion due to the galaxy: sources in the SMBHB's sphere of influence will have faster velocity (cf.~Eq.~\eqref{eq:slow_cond_sphere_infl}), increasing the rates further.

Let us now consider the motion of the central caustic as a result of the orbit of the SMBHB. We will again focus on tight binaries $d\ll 1$ and consider static stars. The rate for an individual system is given by

\begin{equation}
\mathcal{R}_{c1} = \frac{4}{T} n_\star V_{c1}\,,
\end{equation}
where the coefficient reflects the fact that there are 4 rays in the caustic. Because each ray consists of 2 caustic lines that meet at the cusp (Fig.~\ref{fig:caustic}), an event of this type will usually be associated with two magnification peaks.

During an orbit of the SMBHB the central caustic covers a cylinder of radius $R_c$. We will focus on the region corresponding to $d<1$, or lens-source distance above $D_{\rm tight} = a^2/(4R_{\rm g}) = 0.06\,\mathrm{pc}\, M_{10}^{-1/3}T_{\rm yr}^{4/3}$. 
The volume of the cylinder with area $\frac{\pi}{4} \xi_0^2 d^4$
is approximately 
\begin{align}\label{eq:Vc1}
 V_{c1}&\approx
 \frac{\pi}{4} \int_{D_{\rm tight}}^{\tilde D} \mathrm{d}D_{\rm LS} \,\xi_0^2 d^4 = \frac{\pi a^4}{16 R_{\rm g}} \ln \left(\frac{4R_{\rm g} \tilde{D}}{a^2}\right)
\nonumber\\&=
 4.6 \cdot 10^{-5} {\rm pc^3} T_{\rm yr}^{8/3}M_{10}^{1/3} F_1\left(\tilde D,M,T\right)\,,
\end{align}
where $F_1\equiv 1 +
 0.1\ln\left(T_{\rm yr}^{-4/3} M_{10}^{1/3}\tilde D_{\rm kpc}\right)$ for $T<T_{\rm max}$ and zero otherwise.
This expression is only defined for $T<T_{\rm max}=1500 {\,\rm yr}\, M_{10}^{1/4}\tilde D_{\rm kpc}^{3/4}$, corresponding to $d<1$.
The corresponding caustic ray-crossing rate is 
\begin{equation}\label{eq:rate_central_rotating}
    \mathcal{R}_{c1} = 1.9\cdot 10^{-4}  {\rm yr}^{-1} M_{10}^{1/3} T_{\rm yr}^{5/3}\left(\frac{n_\star}{{\rm pc}^3}\right) F_1\left(\tilde D,M,T\right)\,,
\end{equation}
This rate is conservative, as the sources below $D_{\rm tight}$ ($d>1$) have been neglected.
Each event corresponds the crossing of either a cusp or $2\times$ fold lines.

Finally, let us consider the rate for secondary caustic crossing due to the SMBHB orbit. 
Secondary caustics (triangular) span an annulus with radius $R_2 \approx \xi_0/d$ and width $\delta R_2  \approx \xi_0 d^3$ (gray band in Fig.~\ref{fig:caustic}). A point in this ring-like region enters the secondary caustic region twice per orbital period, corresponding to a rate 
\begin{equation}
\mathcal{R}_{c2} = \frac{2}{T} n_\star V_{c2}\,.
\end{equation}
As previously, the crossing event will typically be associated with two magnification peaks. 
The annulus spanned by the revolution of the secondary caustic has an area $2`\pi R_2 \delta R_2 = 2 \pi \xi_0^2 d^2$ if $d\ll 1$, which when integrated over $D_{\rm LS}$
yields a volume 
\begin{align}
    V_{c2} &=  2 \pi\int_{D_{\min}}^{\tilde D} \mathrm{d}D_{\rm LS}\, \xi_0^2 d^2 = 2\pi a^2 (\tilde{D}-D_{\min})
    \nonumber\\
    &\approx  0.69\, {\rm pc}^3 \, T_{\rm yr}^{4/3} \tilde D_{\rm kpc} M_{10}^{2/3}\,.
\end{align}

The secondary caustic rate is then
\begin{equation}\label{eq:rate_secondary_rotating}
    \mathcal{R}_{c2} \approx 1.4 \text{yr}^{-1} \left(\frac{n_\star}{{\rm pc}^{-3}}\right) T_{\rm yr}^{1/3} \tilde D_{\rm kpc} M_{10}^{2/3}\,,
\end{equation}

Efficient magnification requires that the secondary (triangular) caustic is larger than the source's size, $R_{\rm src}\leq \delta R_2 = a^3/\xi_0^2$, i.e. $D_{\rm kpc} \lesssim a^3/(4R_{\rm g}R_{\rm src})=1.2\, T_{\rm yr}^{1/3}M_{10}^{1/12}R_{10}^{1/6}$ and $d\gtrsim (R_{\rm src}/\xi_0)^{1/3}$. The rate of secondary caustic crossing is larger than Eq.~\eqref{eq:rate_central_rotating}. However, the magnification is typically lower and the crossing time shorter than for the primary caustic, with $\mu_{\rm sec}\sim 100$ and $t_{\rm mag,sec}\sim 12$s, Eq.~(\ref{eq:mag_sec},~\ref{eq:caus_cross_sec}).

The rates increase if the modulation shown above can be identified at a distance $R_{\rm det} > R_c$, see App.~\ref{sec:outside_caustic}, Fig.~\ref{fig:far_source_light_curve}.
$\mathcal{R}_0$ increases by a factor $R_{\rm det}/R_c$ and $\mathcal{R}_{c1}$ by $(R_{\rm det}/R_c)^2$. We will address these situations in future work.

\subsection{Average over orientations}

The predictions for individual SMBHBs~(\ref{eq:qpls_prob}, \ref{eq:rate_static}, \ref{eq:rate_central_rotating}, \ref{eq:rate_secondary_rotating}) consider head-on binaries: we will account for the inclination by projecting on randomly rotated orbital planes.
A head-on binary has projected positions $\vec r^{(0)}_\perp = a(C_{\tau},S_{\tau})$, where $C_x\equiv\cos(x), S_x\equiv\sin(x)$ 
and $\tau\equiv t/T$ is the orbital phase. 
A general rotation of the orbital plane rotation turns $\vec r^{(0)}_\perp \to \vec r_\perp = a(C_\alpha C_\beta C_{\tau} - S_\alpha S_{\tau}, C_\alpha S_{\tau} + S_\alpha C_\beta C_{\tau}) $, where $\beta\in[0,\pi],\alpha\in[0,2\pi)$ are the polar and azimuthal angles associated with the rotation.

Because lensing depends on the projection on the plane perpendicular to the lens, one can substitute $d\to d r_\perp/a$ in the expressions for the rates. 
We note that $\mathcal{R}_0, \mathcal{R}_{c2}$ scale as $d^2\propto \tilde r_\perp^2$, while $P_{\rm QPLS}, \mathcal{R}_{c1}$ scales as $d^4\propto \tilde r_\perp^4$.
Hence, the inclination- and-phase average corresponds to the following effective projected semimajor axis, relative to $a$
\begin{equation}\label{eq:rate_projection}
    \bar s_2 \equiv \frac{\langle r_\perp^2\rangle}{a^2} = \frac{2}{3}\,, \quad 
    \bar s_4 \equiv \frac{\langle r_\perp^4\rangle}{a^4} \approx 0.53\,,
\end{equation}
$\bar s_2$ can be easily obtained analytically,
$\bar s_4$ was computed sampling over a large number of random realizations of $\beta,\alpha,\tau$. 
We will use $\bar s_2, \bar s_4$ to account for projections on the head-on rates. 

\subsection{Results from synthetic populations} \label{sec:probability_catalog}

\begin{table*}[]
    \centering
  
    \center \textbf{QPLS} \quad $T_{0} = 10 $yr, $n_\star=1 {{\rm pc}^{-3}}$ (homogeneous)\\[5pt] 
    \setlength{\tabcolsep}{4pt} 
    \renewcommand{\arraystretch}{1.3}

\begin{tabular}{|l|c c c|c c c|c c c|}
\hline
\multirow{2}{*}{Model} & \multicolumn{3}{c|}{$z<0.3$} & \multicolumn{3}{c|}{$z<1$} & \multicolumn{3}{c|}{$z<5$} \\
\cline{2-10}
 & $N_\mathrm{QPLS}$ & $N(<T_0)$ & $n$ [yr$^{-1}$] & $N_\mathrm{QPLS}$ & $N(<T_0)$ & $n$ [yr$^{-1}$] & $N_\mathrm{QPLS}$ & $N(<T_0)$ & $n$ [yr$^{-1}$] \\
\hline
HS-nod-noSN & $46.9$ & $1\cdot 10^{4}$ & $6\cdot 10^{-3}$ & $336$ & $3\cdot 10^{5}$ & $0.28$ & $716$ & $4\cdot 10^{6}$ & $30.6$ \\
\hline
HS-nod-SN-high-accr & $5.52$ & $2\cdot 10^{3}$ & $3\cdot 10^{-3}$ & $100$ & $8\cdot 10^{4}$ & $0.14$ & $207$ & $8\cdot 10^{5}$ & $10.6$ \\
\hline
LS-nod-noSN & $1.22$ & $855$ & $5\cdot 10^{-4}$ & $19.8$ & $2\cdot 10^{4}$ & $0.021$ & $35.8$ & $2\cdot 10^{5}$ & $0.51$ \\
\hline
LS-nod-SN & $3.77$ & $2\cdot 10^{3}$ & $1\cdot 10^{-4}$ & $37.3$ & $5\cdot 10^{4}$ & $4\cdot 10^{-3}$ & $76.3$ & $2\cdot 10^{5}$ & $0.014$ \\
\hline
\end{tabular}
    \caption{\qpls{} detection rates, for binaries with $(1+z)T<T_0=10$ yr and $q>0.5$ under different SMBH population models~\cite{Barausse:2023yrx} and redshift thresholds. Additional columns show the total number of binaries and merger rates with the same selection limits. Note that the light seed models (bottom two rows) are disvafoured by the high-$z$ quasar luminosity function (Fig.~3 in Ref.~\cite{Barausse:2023yrx}).
    \qpls{} results scale with stellar density as as $n_\star/{\rm pc}^{-3}$.
   }
    \label{tab:numbers_QPLS_10yr}

\center \textbf{QPLS} \quad $T_0 = 40 $yr, $n_\star=1 {{\rm pc}^{-3}}$ (homogeneous)\\[5pt] 

\begin{tabular}{|l|c c c|c c c|c c c|}
\hline
\multirow{2}{*}{Model} & \multicolumn{3}{c|}{$z<0.3$} & \multicolumn{3}{c|}{$z<1$} & \multicolumn{3}{c|}{$z<5$} \\
\cline{2-10}
 & $N_\mathrm{QPLS}$ & $N(<T_0)$ & $n$ [yr$^{-1}$] & $N_\mathrm{QPLS}$ & $N(<T_0)$ & $n$ [yr$^{-1}$] & $N_\mathrm{QPLS}$ & $N(<T_0)$ & $n$ [yr$^{-1}$] \\
\hline
HS-nod-noSN & $5\cdot 10^{3}$ & $4\cdot 10^{4}$ & $6\cdot 10^{-3}$ & $5\cdot 10^{4}$ & $7\cdot 10^{5}$ & $0.28$ & $8\cdot 10^{4}$ & $6\cdot 10^{6}$ & $30.6$ \\
\hline
HS-nod-SN-high-accr & $389$ & $4\cdot 10^{3}$ & $3\cdot 10^{-3}$ & $8\cdot 10^{3}$ & $2\cdot 10^{5}$ & $0.14$ & $2\cdot 10^{4}$ & $1\cdot 10^{6}$ & $10.6$ \\
\hline
LS-nod-noSN & $187$ & $2\cdot 10^{3}$ & $5\cdot 10^{-4}$ & $993$ & $3\cdot 10^{4}$ & $0.021$ & $2\cdot 10^{3}$ & $2\cdot 10^{5}$ & $0.51$ \\
\hline
LS-nod-SN & $341$ & $3\cdot 10^{3}$ & $1\cdot 10^{-4}$ & $4\cdot 10^{3}$ & $9\cdot 10^{4}$ & $4\cdot 10^{-3}$ & $5\cdot 10^{3}$ & $3\cdot 10^{5}$ & $0.014$ \\
\hline
\end{tabular}
\caption{Same as Tab.~\ref{tab:numbers_QPLS_10yr} but for longer periods $(1+z)T<T_0=40$yr.}
\label{tab:numbers_QPLS_40yr}
\end{table*}

\begin{table*}[t]
    \centering
    \setlength{\tabcolsep}{4pt} 
    \renewcommand{\arraystretch}{1.3}
    \center \textbf{Caustic crossings $[{\rm yr}^{-1}]$} \quad $a_0=10^3 GM/c^2 q_s^{\frac{16}{37}}$, $n_\star=1 {{\rm pc}^{-3}}$ (homogeneous)\\[5pt] 
\begin{tabular}{|l|c c c c|c c c c|c c c c|}
\hline
\multirow{2}{*}{Model} & \multicolumn{4}{c|}{$z<0.3$} & \multicolumn{4}{c|}{$z<1$} & \multicolumn{4}{c|}{$z<5$} \\
\cline{2-13}
 & $R_0$ & $R_{{c1}}$ & $R_{{c2}}$ & $N(<a_0)$ & $R_0$ & $R_{{c1}}$ & $R_{{c2}}$ & $N(<a_0)$ & $R_0$ & $R_{{c1}}$ & $R_{{c2}}$ & $N(<a_0)$ \\
\hline
HS-nod-noSN & $2\cdot 10^{4}$ & $1\cdot 10^{5}$ & $4\cdot 10^{5}$ & $10\cdot 10^{4}$ & $8\cdot 10^{4}$ & $4\cdot 10^{5}$ & $2\cdot 10^{6}$ & $2\cdot 10^{6}$ & $10\cdot 10^{4}$ & $4\cdot 10^{5}$ & $3\cdot 10^{6}$ & $7\cdot 10^{6}$ \\
\hline
HS-nod-SN-hi-ac & $2\cdot 10^{3}$ & $9\cdot 10^{3}$ & $4\cdot 10^{4}$ & $1\cdot 10^{4}$ & $1\cdot 10^{4}$ & $6\cdot 10^{4}$ & $4\cdot 10^{5}$ & $3\cdot 10^{5}$ & $2\cdot 10^{4}$ & $10\cdot 10^{4}$ & $8\cdot 10^{5}$ & $2\cdot 10^{6}$ \\
\hline
LS-nod-noSN & $147$ & $675$ & $4\cdot 10^{3}$ & $3\cdot 10^{3}$ & $532$ & $2\cdot 10^{3}$ & $2\cdot 10^{4}$ & $4\cdot 10^{4}$ & $1\cdot 10^{3}$ & $4\cdot 10^{3}$ & $5\cdot 10^{4}$ & $2\cdot 10^{5}$ \\
\hline
LS-nod-SN & $93.6$ & $345$ & $5\cdot 10^{3}$ & $5\cdot 10^{3}$ & $677$ & $2\cdot 10^{3}$ & $4\cdot 10^{4}$ & $1\cdot 10^{5}$ & $1\cdot 10^{3}$ & $4\cdot 10^{3}$ & $8\cdot 10^{4}$ & $4\cdot 10^{5}$ \\
\hline
\end{tabular}

    \caption{Total caustic crossing rates [yr$^{-1}$] and number of binaries for different redshift ranges and SMBHB population models~\cite{Barausse:2023yrx}.
Columns correspond, respectively to static caustic crossing, central rotating caustic and secondary (triangular) rotating caustic, assuming $\tilde D_{\rm kpc}=1$, see  Eqs. (\ref{eq:rate_static},\ref{eq:rate_central_rotating},\ref{eq:rate_secondary_rotating}) for scalings.
In addition to the redshift threshold, only sources with $a<10^3 GM$ (GW-driven evolution) and $q>0.5$ are considered.
$N(<a_0)$ shows the total number of binaries obtained with the same selection cuts. 
Note that the light seed models (bottom two rows) are disvafoured by the high-$z$ quasar luminosity function (Fig.~3 in Ref.~\cite{Barausse:2023yrx}). All results scale as $n_\star/{\rm pc}^{-3}$.
}
\label{tab:caustic_rates}

    \centering
    \center \textbf{Caustic crossings $[{\rm yr}^{-1}]$} \quad $a_0=3\cdot 10^3 GM/c^2 q_s^{\frac{16}{37}}$, $n_\star=1 {{\rm pc}^{-3}}$ (homogeneous)\\[5pt] 
    \setlength{\tabcolsep}{4pt} 
\begin{tabular}{|l|c c c c|c c c c|c c c c|}
\hline
\multirow{2}{*}{Model} & \multicolumn{4}{c|}{$z<0.3$} & \multicolumn{4}{c|}{$z<1$} & \multicolumn{4}{c|}{$z<5$} \\
\cline{2-13}
 & $R_0$ & $R_{{c1}}$ & $R_{{c2}}$ & $N(<a_0)$ & $R_0$ & $R_{{c1}}$ & $R_{{c2}}$ & $N(<a_0)$ & $R_0$ & $R_{{c1}}$ & $R_{{c2}}$ & $N(<a_0)$ \\
\hline
HS-nod-noSN & $2\cdot 10^{7}$ & $2\cdot 10^{7}$ & $6\cdot 10^{7}$ & $8\cdot 10^{6}$ & $6\cdot 10^{7}$ & $2\cdot 10^{8}$ & $3\cdot 10^{8}$ & $1\cdot 10^{8}$ & $7\cdot 10^{7}$ & $3\cdot 10^{8}$ & $4\cdot 10^{8}$ & $6\cdot 10^{8}$ \\
\hline
HS-nod-SN-hi-ac & $1\cdot 10^{6}$ & $3\cdot 10^{6}$ & $5\cdot 10^{6}$ & $10\cdot 10^{5}$ & $10\cdot 10^{6}$ & $3\cdot 10^{7}$ & $5\cdot 10^{7}$ & $2\cdot 10^{7}$ & $2\cdot 10^{7}$ & $6\cdot 10^{7}$ & $1\cdot 10^{8}$ & $2\cdot 10^{8}$ \\
\hline
LS-nod-noSN & $1\cdot 10^{5}$ & $3\cdot 10^{5}$ & $6\cdot 10^{5}$ & $2\cdot 10^{5}$ & $4\cdot 10^{5}$ & $2\cdot 10^{6}$ & $3\cdot 10^{6}$ & $3\cdot 10^{6}$ & $8\cdot 10^{5}$ & $3\cdot 10^{6}$ & $7\cdot 10^{6}$ & $2\cdot 10^{7}$ \\
\hline
LS-nod-SN & $7\cdot 10^{4}$ & $4\cdot 10^{5}$ & $7\cdot 10^{5}$ & $4\cdot 10^{5}$ & $5\cdot 10^{5}$ & $2\cdot 10^{6}$ & $6\cdot 10^{6}$ & $9\cdot 10^{6}$ & $8\cdot 10^{5}$ & $3\cdot 10^{6}$ & $1\cdot 10^{7}$ & $3\cdot 10^{7}$ \\
\hline
\end{tabular}

    \caption{Same as Tab.~\ref{tab:caustic_rates} but for quasi-circular binaries with separation $a\leq 3\cdot 10^3 GM q_s^{\frac{16}{37}}$. 
    }
    \label{tab:caustic_rates_2}
\end{table*}

The \qpls{} probability (\ref{eq:qpls_prob_base}) and rates-per-binary (\ref{eq:rate_static},\ref{eq:rate_central_rotating},\ref{eq:rate_secondary_rotating}) can be used to estimate the total number of caustic crossings. To this end, we adopt a set of SMBHB population models based on semi-analytic prescriptions calibrated on cosmological simulations~\cite{Barausse:2023yrx},
which have been compared to pulsar-timing array (PTA) results~\cite{NANOGrav:2023gor,EPTA:2023fyk,Reardon:2023gzh,Xu:2023wog} and high-redshift quasar luminosity~\cite{Shen:2020obl}. These models track black hole seeding, growth via accretion, and mergers following galaxy interactions, with variations in seed mass, supernova feedback, and accretion efficiency (see also Refs.~\cite{Barausse:2012fy,Klein:2015hvg,Barausse:2020mdt,Toubiana:2024bil}). 
We consider four scenarios.
\begin{enumerate}
    \item \textbf{HS-nod-noSN}: This model assumes \emph{heavy seeds} formed from direct collapse at high redshift, and neglects the effects of supernova (SN) feedback on black hole growth. Mergers between black holes are assumed to follow galaxy mergers with minimal delay. We consider the extrapolation to infinite grid resolution to provide the most optimistic case.

    \item \textbf{HS-nod-SN-hi-ac}: Like the previous model, this scenario adopts heavy seeds and short delays between galaxy and black hole mergers. It includes SN feedback (reducing the gas reservoir in low-mass galaxies) but incorporates more efficient accretion. 

    \item \textbf{LS-nod-noSN}: This model is based on \emph{light seeds}, such as remnants of Population III stars, with short time delays.
    It does not include SN feedback, allowing more efficient growth. 

    \item \textbf{LS-nod-SN}: This model combines light seeds and short delays with SN feedback, which suppresses accretion in low-mass galaxies. As a result, black hole growth is quenched in the early universe. 
\end{enumerate}
The heavy seed models are in agreement with both PTA and bright quasars at redshift $z \sim 6$, respectively Figs.~1~\&~3 in Ref.~\cite{Barausse:2023yrx}. Light seeds with negligible SN feedback agree with PTA but fail to reproduce the high-z quasar luminosity function. Light seeds with SN feedback is in disagreement with both datasets, we include it as a pessimistic case.

We  convert the merger rate from synthetic SMBHB catalogs into the number of binaries with orbital frequency above $f_0$ using Eq.~(\ref{eq:dN_df})
\begin{equation}\label{eq:total_binaries}
    N = 4\pi \sum_i \int_{f_{0,i}}^{\infty}\mathrm{d}f\left.\frac{d n}{dt}\right|_i \left.\frac{dt}{df}\right|_i {D_C(z_i)^2}\,.
\end{equation}
Here $i$ labels the catalog samples (including selection criteria, see below), $dn/dt\big|_i$ is number of mergers per comoving volume and time, $D_C$ is the comoving distance, $dt/df\big|_i$ is the time-in-band (Eq.~\ref{eq:dt_df}) and $f_{0,i}$ is the minimum frequency 
(see Eqs.~27-30 in \cite{Phinney2001}; Eqs.~4-6 in \cite{Kocsis_Sesana2011}).
All quantities are evaluated in the source frame, frequencies refer to GWs $f=2/T$, a subscript $i$ indicates dependence on the binary properties ($z_i,M_i,q_i,\cdots$), the sum is equivalent to an integration over redshift and binary properties.%
\footnote{For catalog details see \url{https://people.sissa.it/~barausse/catalogs/models_LISA_after_PTAs/readme.pdf}}

The sum over $i$ includes selection cuts on the population. We impose a mass limit $M>10^5M_\odot$ and explore different redshfit ranges, $z<0.3,1,5$. Because the estimates in App.~\ref{app:qpls_prob},\ref{app:caustic_crossing_rate} assume equal-mass binaries, we consider only systems with mass ratio $q>0.5$.  We note that most of the SMBHBs in the populations have large mass ratios, the number roughly scaling as $N^*_{\rm SMBHB}\propto q_{\rm min}^{-1}$ (for $q\gtrsim 10^{-2}$). Our results are conservative, as asymmetric binaries may contribute substantially to the number of \qpls{} and caustic crossings.

The \textit{total number of \qpls{} sources} is
\begin{equation}\label{eq:total_binaries_QPLS}
    N_{\rm QPLS} = 4\pi\bar s_4 \sum_i \int_{f_{0,i}}^{\infty}\mathrm{d}f\mathcal{N}_{\rm QPLS}\Big|_i \left.\frac{d n}{dt}\right|_i  \left.\frac{dt}{df}\right|_i {D_C(z_i)^2}\,.
\end{equation}
Here $\bar s_4$ accounts for the random orientation of the orbital plane (Eq.~\ref{eq:rate_projection}) and $\mathcal{N}_{\rm QPLS}\big|_i$ depends on the SMBHB population via $M_i$ and the orbital period $T=2/f$, cf.~\eqref{eq:qpls_prob}. We will consider a minimum GW frequency corresponding to $T<T_0=10, 40$ yr and $N_{\rm orb}=1$, corresponding to at least 4/8 cusp/fold crossings. In this limit we can set $F_Q \sim 1$, as $F_Q\sim \mathcal{O}(1)$ when $N_{\rm orb}=1$.
Note that our results are relative to a homogeneous stellar density $n_\star=1 {\rm pc}^{-3}$.

Tables \ref{tab:numbers_QPLS_10yr} and \ref{tab:numbers_QPLS_40yr} show the total number of \qpls{} sources for systems with periods $T<T_0=10/(1+z)$ and $40/(1+z)$ years, respectively, assuming different SMBHB populations and redshift ranges. We have considered only SMBHBs with comparable masses ($q>0.5$), which excludes most systems in the synthetic population (see above). 
For reference, we include the total number of binaries compatible with the selection cuts $N(T<T_0)$ and the merger rate per year $n$.
Even for our conservative stellar density, the number of \qpls{} sources with $T<10$yr exceeds the yearly merger rate when considering low redshifts events, and also for the most pessimistic LS-nod-SN up to $z<5$.
The detectable sources are dominated by the systems with larger separations (i.e. longer orbital periods), as can be seen by comparing Tables \ref{tab:numbers_QPLS_10yr} and \ref{tab:numbers_QPLS_40yr}.

The \textit{total caustic-crossing rates} are then given by
\begin{equation}\label{eq:total_rate}
    R_K = 4\pi\bar s_K \sum_i \int_{f_{0,i}}^{\infty}\mathrm{d}f\left.\frac{\mathcal{R}_{K}}{1+z}\right|_i\left.\frac{d n}{dt}\right|_i \left.\frac{dt}{df}\right|_i {D_C(z_i)^2}\,.
\end{equation}
Here $K\in \{0,c1,c2\}$ labels the rate, the factor $(1+z_i)^{-1}$ accounts for time dilation due to cosmological redshift and $\bar s_K$ is the projection factor: $\bar s_4$ if $K=c1$, $\bar s_2$ if $K\in\{0,c2\}$, see Eq.~\eqref{eq:rate_projection}. 
The individual rate $\mathcal{R}_{K,i}$ depends on $M_i$ and the frequency via the orbital period as $T=2/f$, see Eqs.~(\ref{eq:rate_static}, \ref{eq:rate_central_rotating}, \ref{eq:rate_secondary_rotating}). As before, we will consider only systems with $M>10^5M_\odot $ and $q>0.5$.
sources with frequency above $f_0$.

For our estimate to be valid we also need to restrict the analysis to systems whose orbital evolution is dominated by GW emission, i.e.~choosing $f_0$, so Eq.~(\ref{eq:dt_df}) is valid. Following Ref.~\cite{Haiman_Kocsis_Menou2009}, we set the minimum frequency such that $a<10^3 M q_s^{{16}/{37}}$, where $q_s = 4q/(1+q)^2$ 
is the symmetric mass ratio. This threshold provides an estimate for the transition from GW to viscosity driven evolution due to gas: it ignores binaries with larger separations and is conservative for gas-poor systems.
These choices are conservative: SMBHBs with larger separations will contribute additional caustic crossings.
For reference, we will also quote the total number of SMBHBs fulfilling our selection cuts for each of the population models.

Table \ref{tab:caustic_rates} summarizes total caustic-crossing rates for the different SMBHB formation scenarios and three maximum values of the redshift. 
The rates associated with the orbital motion ($R_{c1}, R_{c2}$) of the binary are larger than stellar crossings ($R_0$), with rotation of the secondary (triangular) caustic being the largest contribution. Note that the rates associated to stellar motion $R_0$ do not include the enhanced velocity dispersion in the SMBHB sphere of influence and are thus conservative.

Our binary selection criteria have a substantial impact on the rates.
In Table \ref{tab:caustic_rates_2} we show how the rates increase by 1-2 orders of magnitude if GW emission dominates to larger separations $a<3\cdot 10^3 q_s^{\frac{16}{37}} GM/c^2$, i.e. Eq.~\eqref{eq:dt_df} remains valid to lower frequencies. This suggests that SMBHBs whose evolution is stalled or governed by other dynamical processes will also contribute significantly.
The fiducial choice of the maximum distance $\tilde D_{\rm kpc}=1$ in $R_0, R_{c,2}$ is conservative, as sources can be found at much larger distances.

All estimates depend strongly on our fiducial choice of stellar density, 
which was set to $n_\star \sim {\rm pc}^{-3}$ as a reference value. 
A more detailed analysis should account for the spatial distribution of stars (e.g. singular isothermal sphere or S\'ersic profile) and the luminosity function, including observational uncertainties.
However, our results can be easily rescaled to reflect fiducial populations. For instance, the density of giant stars in the solar neighborhood is $\sim10^{-4} \, {\rm pc}^{-3}$ \cite{Bovy:2017}, the density of O/B stars in the solar neighborhood is $\sim10^{-5}\, {\rm pc}^{-3}$ \cite{O_Bstars_solar_neighbour}. The density of red clump stars in the Milky Way bulge is $\sim 10^{-3}{\rm pc}^{-3}$~\cite{density_red_clump_stars}. In the central parsec of the Milky Way, more than 6000 bright late-type stars \cite{Genzel_review} and more than 100 young O/B stars \cite{young_stars} are observed.

Future studies need to address \qpls{} and caustic-crossing rates with increasing level of detail, including unequal-mass binaries, the SMBHBs' dynamical evolution, stellar populations, magnification bias, and survey-specific detection thresholds.

\begin{figure}[t!]
    \centering
    \includegraphics[width=0.7\linewidth]{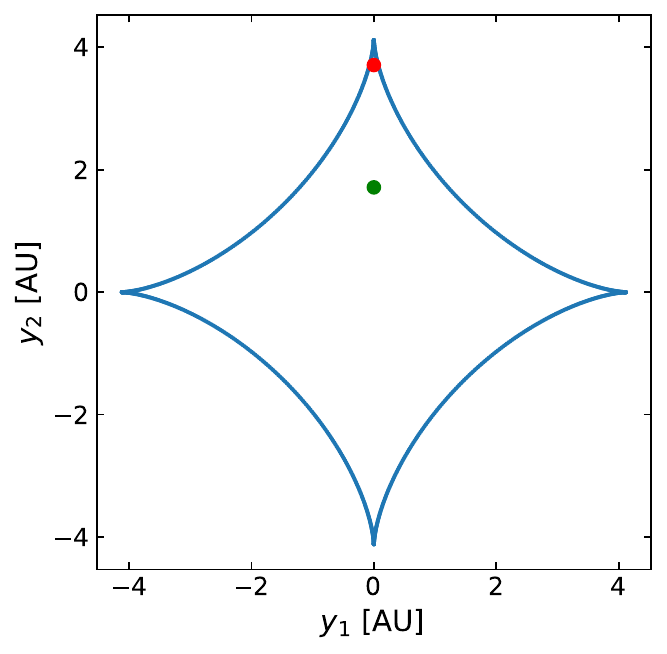}
    {  
    \caption{The caustic curve of a \qpls{} system with two source stars marked by the red and green dot (the size of the stars is not drawn to scale). We assume that the intrinsic luminosity of the 'red' star is twice of that of the 'green' star. Here $M=10^{10}M_{\odot}$, $T = 0.6$ yr, $R_{10}=1$, $e=0$, $D_{\rm LS} =1$ kpc.}\label{fig:two_star_caustic}}
\end{figure}

\begin{figure*}[t]
    \includegraphics[width=1\linewidth]{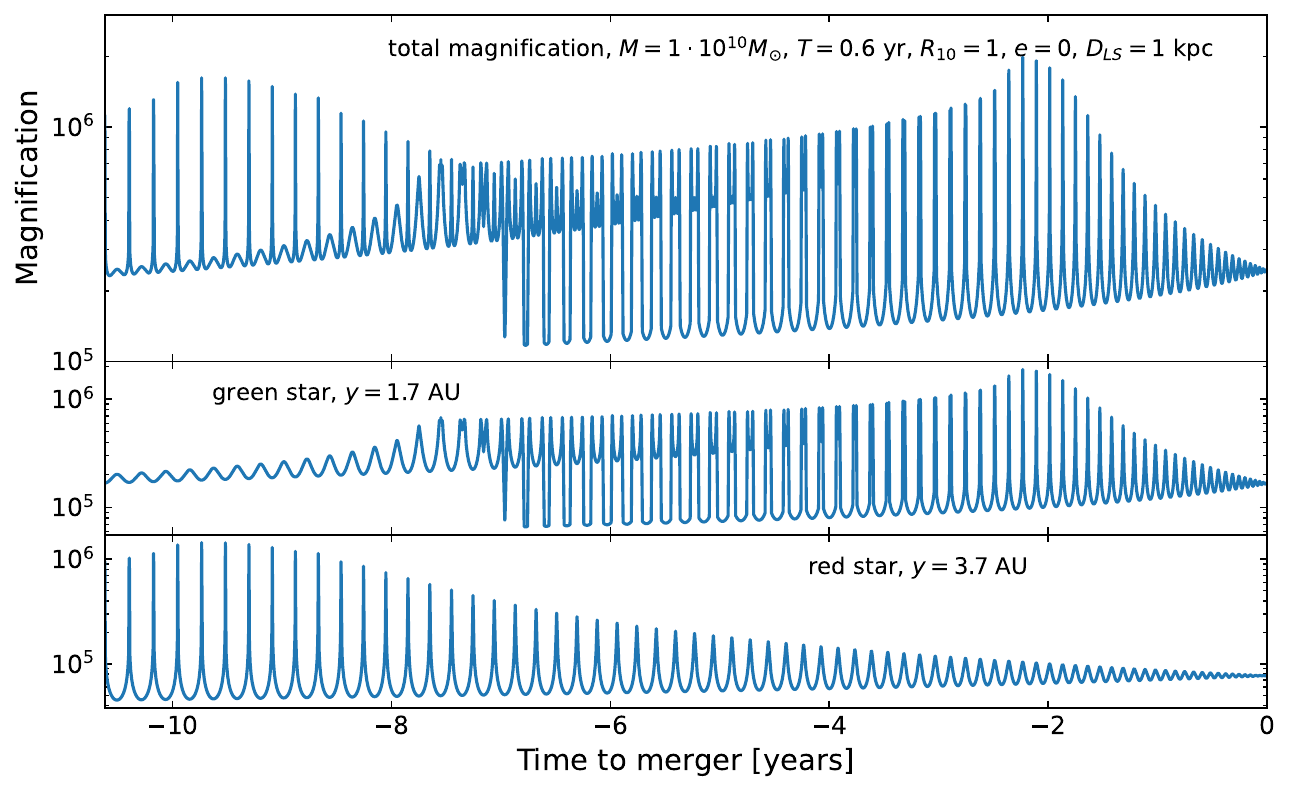}
{  
    \caption{The light curves of the \qpls{} system shown in Fig.~\ref{fig:two_star_caustic}. The middle and bottom panel show the individual light curves of the stars marked as the green and red dots in Fig.~\ref{fig:two_star_caustic} respectively, where the legends also show the distance of the star from the center of mass in the source. The top panel shows the resulting light curve. Note that the proper motion of the stars is neglected here; the magnification diminishes when the sources move out of the caustic region (not shown).}\label{fig:doublestar_lightcurves}}
\end{figure*}

{  
\section{Multiple sources}\label{mult_source}

In this section, we study the scenario when multiple sources undergo \qpls{} by the same SMBHB. Using Eq.~\eqref{eq:qpls_prob}, we approximate the probability of having, e.g. two \qpls{} sources using Poisson statistics:
\begin{align}
    \label{eq:P_two_sourceA}
    P_{\rm QPLS}(2) &\approx \frac12 \mathcal{N}_{\rm QPLS}^2 \notag \\
    &= 8\cdot10^{-10} P_0^2 \left(\frac{n_\star}{{\rm pc}^{-3}}\right)^2 T_{\rm yr}^{16/3} M_{10}^{2/3}F_Q^2\,,
\end{align}
which is small except for binaries with long period. For example, for a binary with $M_{10}=1$ and $T_{\rm yr}=10$, $P_{\rm QPLS}(2)=7\cdot10^{-5}$, which can be significant in surveys with more than $10^4$ galaxies hosting SMBHBs. 

If we adopt the more optimistic estimate of $\mathcal{N}_{\rm QPLS}^{\rm iso}$ in Eq.~\eqref{eq:N_qpls_SIS} assuming a SIS distribution, the probability of having multiple \qpls{} sources will be much higher. If we assume a Poisson statistics as in Eq.~\eqref{eq:P_two_sourceA}, the probability of having two \qpls{} sources for an SIS distribution is 
\begin{align}
    \label{eq:P_two_sourceiso}
    P_{\rm QPLS}^{\rm iso}(2) &\approx\frac12 (\mathcal{N}_{\rm QPLS}^{\rm iso})^2 =0.03\left(\frac{N_{\rm c, pc}}{10^3}\right)^2\, M_{10}^2 P_0^2
\end{align}
Note, however, that these estimates assume that the lensed stars' locations are drawn independently, but in reality the stars positions may be correlated (e.g. if bright stars form in the same region and form binaries or higher multiples), which will lead to higher rates of multi-source QPLS relative to these estimates. However given that $\mathcal{N}_{\rm QPLS}^{\rm iso}\ll 1$ even in the optimistic SIS case, it seems unlikely that a very large number of stars blurs the lightcurve by mimicking an extended homogeneous source.
A more detailed assessment of these correlations will be left for future work.

If multiple \qpls{} sources exist near the caustic, the resulting light curve will be a superposition of  light curves from individual sources. 
Each light curve will have the same periodicity; hence, multiple sources will not affect the inference of the SMBHB period from \qpls{}. We can potentially identify the existence of multiple sources if they cross the caustic curves at their respective source planes at different times. 

An example of double \qpls{} sources is shown in Fig.~\ref{fig:two_star_caustic} and Fig.~\ref{fig:doublestar_lightcurves}. Fig.~\ref{fig:two_star_caustic} shows the caustic curve and the position of the two stars as marked by the red and green dots. We assume that both stars have the same $D_{\rm LS}$ for simplicity. Fig.~\ref{fig:doublestar_lightcurves} shows the individual light curves of the 'green' and 'red' stars in the middle and bottom panel. The top panel shows the resulting light curve of the whole system as a superposition of the two individual light curves, assuming that the intrinsic luminosity of the 'green' star and the 'red' star is the same. The 'red' star is near the cusp of the caustic curve initially hence its light curve displays strong amplitude modulation in earlier times. The amplitude modulation grows weaker as the binary inspirals and the caustic curve shrinks, getting further away from the 'red' star. The 'green' star is much nearer the center of mass than the 'red' star. It only marginally crosses the innermost fold points initially, as indicated by the weak amplitude modulation of its individual light curve in earlier times. In later times as the binary inspirals, the 'green' star starts to cross the fold and eventually the cusp of the caustic curve, displaying stronger amplitude modulation. We thus see the superposition of two periodic magnification patterns in the resulting light curve at different stages of the inspiralling of the SMBHB. This clearly indicates the presence of multiple \qpls{} sources behind the SMBHB.

Note that the two stars that get lensed by the same SMBHB might form a binary. In this case, the two stars will orbit around each other instead of staying relatively stationary as assumed in the previous example. This will produce even more complicated light curves, containing also information about the dynamics of the stars. We will leave the detailed exploration of this case for future study.}

\bibliography{gw_lensing}
\end{document}